**A minimum set of stable blocks for rational design of polypeptide chains**

Running head: **A set of stable blocks for protein rational design**

Alexei N. Nekrasov[1], Ludmila G. Alekseeva[1], Rudolf A. Pogosyan[1], Dmitry A. Dolgikh[1], M.P. Kirpichnikov[1], Alexandre G. de Brevern[2*], Anastasia A. Anashkina[3]

[1] Shemyakin-Ovchinnikov Institute of Bioorganic Chemistry, The Russian Academy of Sciences, Miklukho-Maklaya St. 16/10, 117997, Moscow, Russia
[2] INSERM UMR S-1134, DSIMB, Univ. Paris Diderot, Sorbonne Paris Cite, Univ de la Reunion, Univ des Antilles, INTS, lab. of excellence GR-Ex 6, rue Alexandre Cabanel 75739 Paris Cedex 15, France
[3] Engelhardt Institute of Molecular Biology, Russian Academy of Sciences, Vavilov St. 32, 119991 Moscow, Russia

**Contacts:** alexandre.debrevern@univ-paris-diderot.fr, nastya@eimb.ru

**Supplementary Information**. Supplement.doc

## Abstract

The aim of this work was to find a minimal set of structurally stable pentapeptides, which allows forming a polypeptide chain of a required 3D structure. To search for factors that ensure structural stability of the pentapeptide, we generated peptide sequences with no more than three functional groups, based on the alanine pentapeptide AAAAA. We analyzed 44,860 structures of peptides by the molecular dynamics method and found that 1,225 pentapeptides over 80% of the simulation time were in a stable conformation. Clustering of these conformations revealed 54 topological types of conformationally stable pentapeptides. These conformations relate to different combined elements of the protein secondary structure. So, we obtained a minimal set of amino acid structures of conformationally stable pentapeptides, creating a complete set of different topologies that ensure the formation of pre-folded conformation of protein structures.

## 1. Introduction

The folding of short peptide fragments is seldom discussed as they normally adopt random-coil conformations in water. It is believed that interacting with proteins, peptides take a certain conformation. Different authors underlined that both individual peptides in solution and peptides within proteins can have conformational preferences. They can be detected experimentally by examining peptides in solution, analyzing the structure of proteins, and conducting molecular dynamics in silico. Hatakeyama and co-workers have studied the latent propensity of short peptides to adopt folded conformations, alanine-rich peptides were placed in the cavity of self-assembled host. They found that these peptide fragments adopted specific conformations within the protected cavity [1]. Study of PDB-files demonstrated that of the 160,000 combinatorially possible tetrapeptides 1,500 adopted conformations similar to protein structures [2]. By analyzing 133 8-residue fragments from six different proteins by molecular dynamics, it was found that 85 peptides do not have a preferred structure, and 48 of them converge to the preferred structure [3].

Analyzing the structure of proteins from the Protein DataBank (https://www.rcsb.org/), de Brevern et al. proposed a structural alphabet of small 3D structural prototypes called Protein Blocks (PBs) [4]. This structural alphabet includes 16 PBs, each one is defined by the (phi, psi) dihedral angles of 5 consecutive residues [5]. The size of Protein Blocks corresponds to the maximum length of fragments preserving a minimum of informational entropy [6].

Three-dimensional protein structures can be described with a library of 3D fragments that define a structural alphabet. Etchebest et al. have shown that the geometrical features of the different PBs are preserved (local RMSD value equal to 0.41 Å on average) and sequence-structure specificities reinforced when databanks are enlarged [7]. PBs are short motifs capable of representing most of the local structural features of a protein backbone. Alignment of these local features as sequence of symbols enables fast detection of structural similarities between two proteins [8]. PB-based pairwise structural alignment method based on dynamic programming with a dedicated PB Substitution Matrix (SM) gave an excellent performance, when compared to other established methods for mining [9]. Based on PB, multiple structural alignment method realized in a web server called multiple Protein Block Alignment (mulPBA) [10]. Using the algorithm based on protein blocks, a database of structural alignments (DoSA) was created. DoSA provides unique information about 159,780 conformationally similar and 56,140 conformationally dissimilar Structurally Variability Regions in 74,705 pairwise structural alignments of homologous proteins [11]. PB description of local structures was used to analyze conformations that are preferred sites for structural variations and insertions, among group of related folds [12]. Local structure prediction tool PB-kPRED based on representation of structure of a protein as a string of Protein Blocks and using the structural information from homologues in preference, if available. The method achieved mean accuracies ranging from 40.8% to 66.3% depending on the availability of homologues [13].

Also the structural alphabet was used to predict the loops connecting two repetitive structures [14]. Furthermore, it has been shown that protein structure can be described by a limited set of recurring local structures. In this context, a library composed of 120 overlapping long structural prototypes (LSPs) representing fragments of 11 residues in length and covering all known local protein structures was established [15]. On the basis of LSPs, a novel prediction method that proposes structural candidates in terms of LSPs along a given sequence was developed. This methodology was used to predict protein flexibility [15]. Then three flexibility classes were defined and proposed a method based on the LSP prediction method for predicting flexibility along the sequence. The method is implemented in PredyFlexy web server [16].

In this paper we focus on a minimal set of structurally stable pentapeptides for rational design of polypeptide chain with desirable three-dimensional structure. The pentapeptides were tested using molecular dynamics simulations. Structurally stable pentapeptides are peptides, which were in the same conformation state more than 80% of the simulation time. To search for factors that ensure structural stability of the pentapeptide, we selected peptide sequences with no more than three functional groups, based on the alanine pentapeptide AAAAA.

## 2. Methods

*2.1 Pentapeptide sequence dataset*

There are $20^5$=3 200 000 possible pentapeptide sequences of 20 types of amino acid residues. Study of stability for this number of pentapeptides by molecular dynamics method is very time consuming. We assumed that the three functional groups should be sufficient to form a stable conformational state of the pentapeptide. We wanted to study the effect of the amino acid residue positions on its stability. So we proposed the following scheme for the generation of pentapeptide sequences.

The alanine pentapeptide was taken as a matrix, since glycines in the polypeptide chain possess additional conformational space due to the absence of a side radical. Three residues of this alanine pentapeptide — one in the central position (position "0") and two more in other positions (positions -2, -1, + 1, and +2) were sequentially replaced by all possible amino acid residues of 20 canonical amino acids. The number of possible pairs of two replaced amino acids positions in the pentapeptide is 6 (substitutions at -2 and -1, -1 and 1, -2 and 2, -1 and 1, -1 and 2, 1 and 2 positions are possible), therefore the number of possible combinations is $20^3 * 6 = 48000$. However, in this approach 20 * (20 * 6 + 6 * 6 + 1) identities are formed, associated with the presence of an alanine matrix. Thus, subtracting the identities, we get 44860 possible unique sequences. It can be seen that there is a strong overrepresentation of alanines at positions -2, -1, +1, +2. All residues except alanine are found at these positions 1160 times each and alanine is found 22820 times. At position 0 all amino acid residues are found 2243 times each.

Due to this scheme for pentapeptide sequence generation one can evaluate the contribution of each position to the pentapeptide stability and the influence of the amino acid type in a particular position on the stability of the pentapeptide.

*2.2 Molecular modeling of the pentapeptide structure by molecular dynamics method*

C-terminus of each pentapeptide was protected by methyl group and N-terminus was protected by acetyl group. These caps have removed the effect of the end charges of the polypeptide chain on the conformational states obtained in MD modeling.

Force field AMBER/OPLS was used [17,18] for molecular dynamics simulation performed with X-PLOR (Version 3.1) software

(http://www.csb.yale.edu/userguides/datamanip/xplor/xplorman/node1.html) with personal license from Axel T. Brünger. The starting conformation of all pentapeptides was unfolded (extended) (φ=180.0º; ψ=180.0º). Molecular modeling of pentapeptides started from extended conformation within 10,000 picoseconds at 300K with 0.001 picosecond integration step. We saved 5,000 current conformational states for each pentapeptide after 5,000 picoseconds of relaxation process, one state per picosecond. This AMBER/OPLS force field was chosen because the original peptide conformation is unfolded. CHARMM force fields work well near local minima, but in the case of three-dimensional hindrances they lead to unpredictable results. We did not use a solvent in the molecular dynamics experiment. Solvent is important for polypeptide chains capable of forming a "hydrophobic core". Pentapeptides are too short to possess this ability. This approach was approved by many years of conformational analysis of short peptides [19,20]. The complete molecular dynamics protocol is provided in the Supplement.

*2.3 Clustering of pentapeptide conformational states*

Molecular dynamics trajectory with 5,000 conformational states was subjected to clustering analysis to split it into a few clusters of the structurally similar conformations. The squared Euclidean distance of torsion angles phi and psi of polypeptide backbone (a total of 8 angles for each pentapeptide) was chosen as distance measure between conformational states of molecular dynamics trajectory. Clustering was carried out by unweighted pair-group method using the centroid average [21] (program "condense", https://fap.sbras.ru). A clustering threshold of 0.15 was used.

Figure 1 shows the size of the two largest clusters of conformational states of molecular-dynamic trajectories for 44,860 pentapeptides. There is a local maximum of the distribution in the lower right corner. It corresponds to the size of the largest cluster of more than 4,000 states, i.e. 80% of the conformational states under consideration. So, the threshold of 80% of the conformational states was chosen for the criterion of structural stability. Pentapeptides where more then 80% of all conformational states were within the same cluster, were called "structurally stable elements" or "structurally stable pentapeptides". The structure closest to the cluster center was selected as the cluster representative. Each structurally stable pentapeptide was characterized by the representative.

*2.4 Clustering of different structurally stable pentapeptides*

To identify common topological types among the obtained structurally stable pentapeptides, representative structures were subjected to the same clustering procedure described above. As a result, 54 clusters of structurally stable structural elements with different topologies were obtained (Supplement, Table 1). Each of 54 cluster of structurally stable pentapeptide was characterized by the structure closest to the cluster center (Supplement, Table 2).

*2.5 Comparison of idealized secondary structures with spatial structures of model pentapeptides*

To identify the functional role that structurally stable pentapeptides can play in the protein structure, a comparison of the Cartesian coordinates of the Cα atoms of three consecutive amino acid residues of the pentapeptide located at the N- and C-termini of the pentapeptides with idealized secondary structures was performed: α-helix ($\varphi$=-57.0º; $\psi$=-47.0º), anti-parallel β-structure ($\varphi$=-139.0º;$\psi$=+135.0º) and parallel β -structure ($\varphi$=-119.0º; $\psi$=+113.0º) [22]. To combine spatial structures, the MaxCluster program was used (http://www.sbg.bio.ic.ac.uk/maxcluster/).

## 3. Results

We supposed that stability could be provided by the interaction of two or three functional groups within the pentapeptide. So, it is no need to consider all 3,200,000 possible sequences of pentapeptides to study the factors of high structural stability of pentapeptides and to identify all possible types of stable conformations. Later we shall prove the validity of this assumption.

Alanine pentapeptide AAAAA has no stable conformation since its lifetime in the most stable conformation was 2.3% of the molecular dynamics simulation time. It can be explained by a short side chain and weak interactions between them. We did not choose glycine pentapeptide GGGGG because glycine has additional conformational space of backbone [22]. In the AAAAA sequence, the residues in the central and in two other positions were consequentially replaced with all possible canonical amino acid residues. After elimination of identical sequences, we obtained 44,860 unique sequences of pentapeptides and carried out molecular dynamics simulation for them (see Methods).

The size distribution of the two largest clusters of molecular-dynamic trajectories for 44,860 pentapeptides is shown in Figure 1. A large peak is observed in the bottom-left corner. In the bottom-right corner, there is a small peak. The local maximum of the distribution in the lower right corner corresponds to the size of the largest cluster 1 of more than 4,000 states, i.e. 80% of the conformational states under consideration (see Supplementary Table 2 for the list of pentapeptides with lifetime in stable conformation over 80%). So, the threshold of 80% of the conformational states was chosen for the criterion of structural stability. Clustering of molecular-dynamic trajectories for 44,860 pentapeptides revealed that in 1,225 pentapeptides of the largest cluster contained more than 80% of the conformational states. Thus, in the examined set of 44,860 pentapeptides, only 2.73% of them were conformationally stable. The most stable peptide was IATAE with stable conformational state 99.56% of the simulation time.

A structure and sequence of central conformation in the largest cluster was chosen for each structurally stable pentapeptide (see Materials and Methods). Further, the central conformation of stable pentapeptides was clustered using the torsion angles of the polypeptide backbone φ and ψ resulting in 54 groups (Supplementary Table 1). The largest group contains 324 stable pentapeptides. A representative structure closest to the center of the cluster was chosen for each cluster of similar 3D structures (see representative structures and their parameters in Supplementary Table 2).

Conformationally stable pentapeptides (KDKAA, PAHVA, AAWCD, RARAY, KNGAA, PAKKA, WAEAW, KREAA, DAICA, RAMAD, EAREA, KGYAA, WAEYA, EAKAK, EKDAA, APKKA, AAPKE, APKGA, PAGAP, FDGAA, PACAK, KAKDA, EAGKA, APPAP, PDPAA, RAEAD) did not join any clusters and formed their own groups with a unique topology of the polypeptide backbone.

For description of the features of the amino acid sequences of structurally stable peptides the central residue in the pentapeptide was named as "0", and for the others "-2" (N terminus), "-1", "+1" and "+2" (C terminus) numeration was used. The balance of ALA will not be discussed in most cases, since it is known that its amount is significantly increased in all positions.

In particular, all 322 sequences of stable peptides that fall into cluster 1 contain at position "+2" 97.8% LYS and 2.2% ARG (Table 1, also see Supplementary_inf.xls file for details). A similar analysis carried out for cluster 2 containing 218 pentapeptides shows in 100% of cases localization at position "+2" of a residue with a negatively charged side group (ASP 78.9% and GLU 21.1%). Negatively charged residues ASP and GLU can be found at the position "-2" in the cluster 3 sequences with the probability 72.3% and 26.1% respectively. Analysis of occurrence of amino acid residues in different positions of pentapeptides for other clusters was not possible due to the small number of peptides (total 107 pentapeptides for clusters 14-54). In general, it can be seen that for each cluster of conformationally stable pentapeptides there is a characteristic pattern of the location of charged amino acid residues (Table 1).

We assumed that the possible functions of structurally rigid pentapeptides can be: the initiation of the formation of elements of the secondary structure, the maintenance of the conformation of the secondary structure, and the termination of the elements of the secondary structure. To identify structurally rigid elements that can play this role, coordinates of three N- or C-terminal C-alpha atoms were compared with idealized elements of the secondary structure. To construct idealized elements of the secondary structure, the following values of the torsion angles of the polypeptide backbone were used for α-helix ($\varphi$ =-57.0º; $\psi$=-47.0º), anti-parallel β-structure ($\varphi$=-139.0º; $\psi$ =+135.0º) and parallel β-structure ($\varphi$ =-119.0º; $\psi$=+113.0º)) [22].

Supplement Table 3 shows the results of comparison of the structures of the terminal regions of rigid pentapeptides with C-alpha atoms of idealized secondary structures. In the 40 of 54 clusters the best value of RMSD were less than 0.1. In 13 clusters, the distances to the nearest idealized structures were in the range from 0.1 to 0.2. For one cluster (namely cluster 43), this distance was 0.207.

The good structural correspondence of stable pentapeptides to elements of the secondary structure of proteins is intriguing and suggests that they can play an important role in the formation of pre-folded conformation of natural polypeptide chains, by initiating and terminating conformations corresponding to the elements of the secondary structure. Pentapeptides, the beginning and end of which correspond well to the same type of secondary structure, should be considered as peptides that maintain this type of secondary structure (Table 2). In particular, pentapeptides from clusters 3, 7, 11, 12, 13, 15, 25, 27, 28, 34, 35, 37, 38, 43, 45, 48, 51 and 52 should be considered as pentapeptides maintaining α-helix. Pentapeptides from clusters 4, 8, 21, 24, 30, 32, 49 and 54 maintain the ↑↑β-structure conformation. We did not found structurally stable peptides that maintain the ↓↑β-structure conformation.

Table 2 roughly summarizes the behavior of each cluster as follows:

- Peptides from clusters 1, 22, 31, 44, 46 and 50 can link α-helix with ↑↓β-structure. Peptides from clusters 10, 16, 20, 23, 39, 42 and 47 forming the link from the ↑↓β -structure to the α-helix.
- Pentapeptides from the clusters 1, 2, 5, 9, 14, 18, 29, 31, 36, 41, 44, 46, 50 and 53 can make a link from α-helix to ↑↑β-structure. A link from ↑↑β -structure to α-helix can be realized by the peptides from the clusters 10, 16, 17, 19, 39, 40 and 42.
- The transition from the ↑↓β-structure to the ↑↑β-structure can be realized by the pentapeptides from the clusters 30 and 33, the pentapeptides realizing the transition from the ↑↑β structure to the ↑↓β -structure can be found in clusters 6 and 54.

The identified structurally stable pentapeptides probably form an almost complete set of topological types of peptides necessary for the formation of the prefolding state of the polypeptide chain. The replacement of the frequently used alanine residues with residues with other functional groups can substantially replenish a set of structurally stable elements, but it is unlikely to change the set of topological types found.

To understand how often conformationally stable peptides are found in native proteins, we analyzed the sequence of human Na, K-ATPase (Uniprot code P05024). The sequence of Na, K-ATPase contains 1,017 overlapping pentapeptides. Molecular modeling showed that 79 structurally stable pentapeptides were identified among them, which cover 32.4% of its sequence.

## 4. Discussion

In this paper, in the framework of our approach [6,23], we proposed a method for searching for conformationally stable pentapeptides for protein folding. For each protein sequence, it is possible to determine conformationally stable sites of the sequence and to predict their spatial structure. Such regions in the protein structure ensure the formation of a prefolding conformational state. A "correct" pre-folding state leads to a decrease in the dimensionality of the multidimensional phase space. A simple gradient descent in such a space can quickly and unambiguously lead to natural polypeptide chains in the native spatial organization. Identification of such sites in proteins allows us to investigate the mechanism of protein folding and to make rational design of native protein sequences.

The performed research revealed topologically complete set of structures that ensures the formation of prefolding conformations for proteins of various topological types. It is important to note that among a limited number of identified types of structurally stable elements there are not only those that initiate formation or maintenance of a certain type of secondary structure, but also terminate them, ensuring the continuation of the way of the polypeptide chain in a definite solid angle.

It is of fundamental importance that 54 topological types of structurally stable pentapeptides were obtained by using in their sequence only three residues with effectively interacting functional groups. The remaining positions in the sequences of structurally stable elements occupied by ALA residues in our study may be occupied by residues that provide interactions in the later stages of folding.

It can be assumed that in real proteins, structurally stable regions can occupy a significant portion of the amino acid sequence of the protein. For example, the sequence of Na, K-ATPase (code Uniprot P05024) has 79 structurally stable pentapeptides, which cover 32.4% of its sequence. It is interesting to note that only 2.73% of the 44 860 pentapeptides studied were structurally stable.

Of further interest is the study of pentapeptides that have two conformational states and that can thus serve as “switches” at the junctions of the moving parts of proteins. Also interesting is the content of conformationally labile pentapeptides capable of adopting the structure dictated by the local environment and the course of the polypeptide chain. In addition, it seems an interesting prospect to study the mutual arrangement of pentapeptides with different conformational properties.

**Acknowledgements**

This study was partially supported by Russian Foundation for Basic Research (#18-54-00037) and was supported by the Program for Molecular and Cellular Biology of the Russian Academy of Sciences. The Na,K-ATPase peptides stability calculation was funded by the Russian Science Foundation (grant #14-14-01152). AdB's work was supported by grants from the Ministry of Research (France), University Paris Diderot, Sorbonne, Paris Cité (France), National Institute for Blood Transfusion (INTS, France), National Institute for Health and Medical Research (INSERM, France) and labex GR-Ex. The labex GR-Ex, reference ANR-11-LABX-0051 is funded by the program "Investissements d'avenir" of the French National Research Agency, reference ANR-11-IDEX-0005-02. AdB acknowledges to Indo-French Centre for the Promotion of Advanced Research / CEFIPRA for collaborative grant (number 5302-2).

We would also like to thanks Axel T. Brünger for the X-PLOR software license. The authors are grateful to Dr. Alexei Adzhubei for attentive reading and valuable comments.

**Legends**

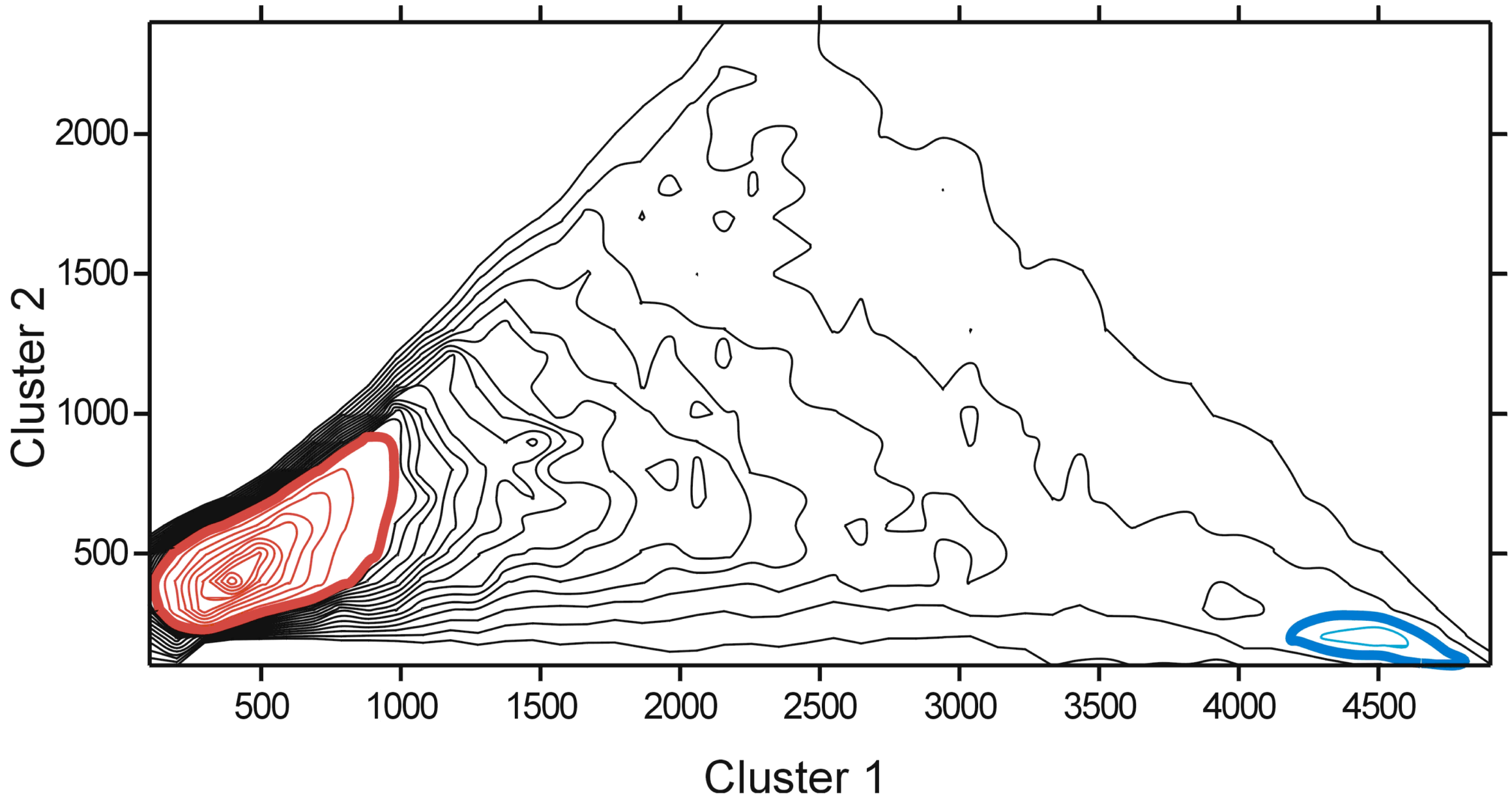

**Figure 1.** The size distribution of the two largest clusters of conformation states of molecular-dynamic trajectories for 44,860 pentapeptides. Structurally unstable peptides are located on the inside of the red line in the bottom-left corner (bold red isoline 300, thin 100). Structurally stable peptides are located inside of the blue line in the bottom-right corner (bold blue isoline 40, thin 20).

Table 1. Sequence patterns of conformationally stable pentapeptides of clusters 1-13. X is any amino acid residue.

| Cluster | Number of sequences | Representative amino acid residue | | | | |
|---|---|---|---|---|---|---|
| | | -2 | -1 | 0 | 1 | 2 |
| 1 | 324 | A/X | A/X | X | A/X | K/R |
| 2 | 216 | A/X | A | X | A | D/E |
| 3 | 186 | D/E | A/K/R | E/D/W/Y | A | A |
| 4 | 92 | R/K | A/K/R | X | A/E/D | A/K/R |
| 5 | 86 | N/A/K/R | A | X | A | E/D |
| 6 | 41 | R/K | E/D | X | A | A |
| 7 | 38 | K/R | A/X | G/X | A | A |
| 8 | 29 | A | A | P | A/X | K/R |
| 9 | 28 | D/E | A | X | A | P |
| 10 | 28 | K/R | A/D/E | P | A | A |
| 11 | 18 | D/E | A | X | A | R/K |
| 12 | 17 | K/R | A/P | X | A/R | A |
| 13 | 15 | D/E | X | X | A | A |
| 14-54 | 107 | A/P/K/R/E/D | A/P/D/E | R/K/E/D/G/P | A/K/R | A/D/E/K/R/P |

Table 2. Clusters of conformationally stable pentapeptides that maintain the elements of the secondary structure and provide a transition from one type of secondary structure to another.

| | | N-termini of pentapeptide | | |
|---|---|---|---|---|
| | | Terminate the α-helix | Terminate the ↑↓β-structure | Terminate the ↑↑β-structure |
| C-termini of pentapeptide | Initiate the α-helix | 3, 7, 11, 12, 13, 15, 25, 27, 28, 34, 35, 37, 38, 43, 45, 48, 51, 52 | 10, 16, 20, 23, 39, 42, 47 | 10, 16, 17, 19, 39, 40, 42 |
| | Initiate the ↑↓β-structure | 1, 22, 31, 44, 46, 50 | - | 6, 54 |
| | Initiate the ↑↑β-structure | 1, 2, 5, 9, 14, 18, 29, 31, 36, 41, 44, 46, 50, 53 | 30, 33 | 4, 8, 21, 24, 30, 32, 49, 54 |

**Supplement. Molecular dynamics protocol.**

A fragment of the command file of the X-PLOR program, describing the initial minimization of the starting conformation, and the MD protocol for the simulation of pentapeptides are given below.

```
  minimize powell
     nstep=500                    {  THE NUMBER OF MINIMIZATION STEP  }
     nprint=1
     drop=10.
  end

{  CONTROL CONFORMATION & ENERGY TERMS  }

        write coordinates output=test.pdb end
        system

  display $zero $ENER $VDW $ELEC $DIHE

{  END OF OUTPUT CONTROL  }

 print threshold=0.1 bonds
 print threshold=10.0 angles
 print threshold=0.0 cdihedrals

{  MOLECULAR DYNAMICS CYCLE  }

    set seed=432324368 end

    vector do (vx=maxwell(300.)) (all)
    vector do (vy=maxwell(300.)) (all)
    vector do (vz=maxwell(300.)) (all)

      vector do (fbeta=100.) (all)

  evaluate ($nucycl=10000)
  evaluate ($time_step=1)

  while($nucycl>0) loop md_cycle_1

{  MOLECULAR DYNAMICS  }

 dynamics verlet
    nstep=1000
    time=0.001
    nprint=999
    ntrfrq=200
    tbath=300.
    tcoupling=true
    firsttemp=300.
    finaltemp=300.
    iasvelocity=maxwell
```

```
  end

{  CONTROL CONFORMATION & ENERGY TERMS  }

  write coordinates output=test.pdb end { Write out coordinates }
  system

  evaluate($num_print=$num_print+1)
  display $num_print $ENER $VDW $ELEC $DIHE

{  END OF OUTPUT CONTROL  }

  evaluate($nucycl=$nucycl-$time_step)

end loop md_cycle_1
```

**Supplementary Table 1.** Sequences of pentapeptides for each of 54 structural clusters formed by pentapeptides with stable spatial conformation.

| Cluster number | Sequences | Conformational stability by molecular dynamics simulation trajectories, (% of the conformational states) |
|---|---|---|
| **Cluster 1** | | |
| | AAMVK | 81.8 |
| | AVVAK | 88.8 |
| | ASCAK | 86.4 |
| | AAFFK | 93.2 |
| | AALFK | 90.1 |
| | FAYAK | 89.9 |
| | AAYYK | 91.9 |
| | ALSAK | 83.5 |
| | AAFLK | 89.3 |
| | AHQAK | 80.4 |
| | YAIAK | 86.2 |
| | AQYAK | 85.1 |
| | AVNAK | 84.8 |
| | TAWAK | 87.2 |
| | AWQAK | 84.1 |
| | AALWK | 89.2 |
| | LAWAK | 88.8 |
| | VAVAK | 90.5 |
| | AALLK | 82.6 |
| | ALHAK | 84.1 |
| | AHFAK | 87.8 |
| | LAAAK | 87.5 |
| | LAFAK | 90.8 |
| | AALAK | 87.7 |
| | AACSK | 83.2 |
| | AAWKK | 85.0 |
| | VATAK | 80.7 |
| | LAIAK | 80.1 |
| | AAWHK | 87.7 |
| | AFMAK | 90.9 |
| | AVIAK | 85.6 |
| | FAVAK | 83.4 |
| | AAHMK | 88.7 |
| | AAMYK | 89.9 |
| | AIYAK | 89.1 |
| | AYYAK | 84.9 |
| | AWAAK | 95.0 |
| | FASAK | 83.4 |
| | AANWK | 87.5 |
| | AMVAK | 83.7 |
| | FATAK | 88.1 |
| | HAHAK | 82.6 |
| | AACWK | 86.8 |
| | AAIAK | 80.0 |
| | TAHAK | 92.0 |
| | AMLAK | 87.5 |
| | AAVAK | 87.6 |
| | AAIHK | 91.6 |
| | ANKAK | 92.2 |
| | AWHAK | 90.4 |
| | AQSAK | 82.8 |
| | AFVAK | 90.6 |
| | AAMFK | 88.4 |
| | FAHAK | 89.7 |
| | AAVLK | 82.7 |
| | LASAK | 87.0 |
| | ACTAK | 86.7 |
| | AAWAK | 81.2 |
| | AFTAK | 81.5 |
| | ATSAK | 81.6 |
| | AAVNK | 83.7 |
| | AYMAK | 84.3 |
| | ATHAK | 92.1 |
| | AMTAK | 85.1 |
| | ACHAK | 85.9 |
| | AALHK | 84.4 |
| | ATVAK | 81.0 |
| | AFSAK | 88.9 |
| | AFIAK | 93.8 |
| | YAVAK | 81.7 |
| | AAIYK | 85.6 |
| | AAVIK | 83.3 |
| | LAVAK | 88.7 |
| | AAQRK | 81.5 |
| | ACVAK | 88.3 |
| | AAAFK | 92.4 |
| | ALLAK | 80.0 |
| | LAYAK | 87.5 |
| | AVWAK | 89.8 |
| | AYHAK | 85.4 |
| | AIFAK | 97.4 |
| | MAAAK | 82.2 |
| | AAYFK | 88.9 |
| | AFFAK | 91.6 |
| | AAHWK | 88.3 |
| | AATIK | 87.8 |
| | MAMAK | 81.9 |
| | AFQAK | 82.8 |
| | ALCAK | 84.2 |
| | AAYLK | 95.1 |
| | AWVAK | 91.6 |
| | AVAAK | 88.9 |
| | TASAK | 88.9 |
| | ACIAK | 89.0 |
| | ALVAK | 81.1 |
| | AAISK | 90.0 |
| | ALTAK | 81.1 |
| | AATCK | 80.9 |
| | AASYK | 81.7 |
| | AVTAK | 91.0 |
| | AWFAK | 95.9 |
| | YASAK | 81.1 |
| | TAMAK | 86.8 |
| | VALAK | 85.8 |
| | FALAK | 90.2 |
| | AAMWK | 82.2 |
| | ASLAK | 87.8 |
| | LATAK | 81.4 |
| | AASFK | 88.7 |
| | ACCAK | 82.2 |
| | WALAK | 86.7 |
| | YAWAK | 90.3 |
| | CAVAK | 89.9 |
| | FAFAK | 89.9 |
| | YAMAK | 89.4 |
| | CAHAK | 83.9 |
| | VASAK | 90.4 |
| | YAQAK | 89.2 |
| | SAMAK | 85.5 |
| | ATLAK | 90.1 |
| | YALAK | 92.6 |
| | MAFAK | 81.1 |
| | IAIAK | 94.7 |
| | WAHAK | 81.7 |
| | AAQWK | 82.5 |
| | ASFAK | 90.9 |
| | WAIAK | 82.6 |
| | IAYAK | 91.5 |
| | TAQAK | 89.7 |
| | LALAK | 85.0 |
| | IAWAK | 88.9 |
| | AFLAK | 88.2 |
| | YAAAK | 88.4 |
| | WAVAK | 82.7 |
| | VAMAK | 81.3 |
| | AATWK | 87.7 |
| | LAMAK | 87.1 |
| | AALYK | 89.2 |
| | AMFAK | 89.8 |
| | SALAK | 81.7 |
| | AWWAK | 80.2 |
| | AYLAK | 94.6 |
| | ACWAK | 86.1 |
| | ATTAK | 92.6 |
| | AANHK | 80.5 |
| | MAYAK | 87.5 |
| | VAFAK | 88.4 |
| | AAQCK | 84.9 |
| | ASYAK | 91.6 |
| | AATYK | 81.1 |
| | AAIMK | 90.5 |
| | ASSAK | 85.1 |
| | ASVAK | 89.8 |
| | AAILK | 87.8 |
| | AACVK | 86.9 |
| | ACQAK | 89.5 |
| | AATAK | 82.8 |
| | AATLK | 88.6 |
| | AIIAK | 95.5 |
| | AAWLK | 95.0 |
| | AAFIK | 93.3 |
| | AAICK | 83.3 |
| | ALWAK | 89.3 |
| | AARQK | 90.9 |
| | AAMLK | 90.7 |
| | ACLAK | 86.9 |
| | AAQVK | 85.8 |
| | AAIIK | 84.8 |
| | ASHAK | 93.3 |
| | AALCK | 89.0 |
| | AAWVK | 92.7 |
| | AAAYK | 82.6 |
| | AFYAK | 91.5 |
| | ACSAK | 85.0 |
| | ACYAK | 85.8 |
| | AAWYK | 80.3 |
| | AAYWK | 88.5 |
| | AAFVK | 80.2 |
| | AAYAK | 89.2 |
| | AFHAK | 84.1 |
| | AVMAK | 88.7 |
| | AYWAK | 94.2 |
| | AWLAK | 90.3 |
| | FAWAK | 91.3 |
| | AATNK | 80.6 |
| | AIAAK | 84.0 |
| | AVLAK | 88.0 |
| | AAYCK | 81.7 |
| | AAACK | 81.2 |
| | AAYHK | 84.6 |
| | AWNAK | 86.1 |
| | AAHCK | 84.1 |
| | AAFAK | 91.2 |
| | AAQLK | 84.8 |
| | AILAK | 95.2 |
| | AMMAK | 80.0 |
| | AAIFK | 90.5 |
| | AAFMK | 92.8 |
| | AAFYK | 89.0 |
| | AAVQK | 93.7 |
| | GAIAK | 83.6 |
| | YAHAK | 84.7 |
| | AIWAK | 86.4 |
| | AAWSK | 83.1 |
| | VAAAK | 87.0 |
| | AAHYK | 95.5 |
| | AAVWK | 88.3 |
| | SAYAK | 84.6 |
| | AVYAK | 89.3 |
| | SASAK | 84.8 |
| | SAVAK | 87.8 |
| | WACAK | 87.7 |
| | VAIAK | 88.4 |
| | HAVAK | 91.7 |
| | ALAAK | 90.5 |
| | AISAK | 92.0 |
| | AAWQK | 87.8 |
| | AAIRK | 81.1 |
| | AAFHK | 89.9 |
| | IAAAK | 84.4 |
| | WANAK | 86.3 |
| | ANHAK | 83.2 |
| | ASTAK | 88.7 |
| | ALMAK | 82.1 |
| | SATAK | 85.4 |
| | ANFAK | 80.4 |
| | AAVHK | 83.4 |
| | AYTAK | 85.7 |
| | AICAK | 90.4 |
| | AITAK | 90.6 |
| | IANAK | 80.3 |
| | AFAAK | 92.3 |
| | AHTAK | 81.6 |
| | WAAAK | 86.5 |
| | AYIAK | 88.2 |
| | CAFAK | 90.6 |
| | AAMHK | 84.9 |
| | AACFK | 80.2 |
| | FAAAK | 88.5 |
| | AAWWK | 83.9 |
| | AAFSK | 86.9 |
| | AANTK | 88.7 |
| | AAYVK | 84.1 |
| | AAMSK | 85.2 |
| | AAFTK | 87.5 |
| | KACAK | 84.4 |
| | TAFAK | 80.2 |
| | AAWMK | 89.7 |
| | AALMK | 82.0 |
| | TALAK | 83.2 |
| | AAALK | 86.8 |
| | TAYAK | 83.8 |
| | RAAAK | 86.9 |
| | AAVTK | 83.6 |
| | ACFAK | 90.7 |
| | AAIQK | 88.3 |
| | AAARK | 80.0 |
| | QAYAK | 86.9 |
| | AAVMK | 83.0 |
| | AATFK | 88.8 |
| | AAQFK | 89.0 |
| | AAHTK | 85.1 |
| | AAIVK | 87.1 |
| | AAINK | 83.7 |
| | AACTK | 89.3 |
| | WAQAK | 80.9 |
| | AAVRK | 84.8 |
| | AAIKK | 87.4 |
| | AAYQK | 85.6 |
| | AAITK | 83.4 |
| | AALRK | 87.5 |
| | AASKK | 83.1 |
| | AAMTK | 83.7 |
| | AACKK | 80.4 |
| | AARKK | 84.0 |
| | AACRK | 86.9 |
| | RAMAK | 95.6 |
| | AAYRK | 83.2 |
| | AAHKK | 86.7 |
| | AAYMK | 84.5 |
| | AASMK | 80.1 |
| | AAHFK | 85.5 |
| | AATKK | 81.1 |
| | AALKK | 88.0 |
| | AYSAK | 81.1 |
| | AAVYK | 89.8 |
| | AAAQK | 83.8 |
| | ATYAK | 87.7 |
| | WATAK | 80.8 |
| | WAFAK | 87.6 |
| | AARLK | 90.1 |
| | IASAK | 94.6 |
| | SAFAK | 84.3 |
| | IALAK | 87.8 |
| | AFWAK | 82.9 |
| | SAAAK | 83.9 |
| | MAVAK | 83.3 |
| | CATAK | 81.4 |
| | AALSK | 81.3 |
| | AAHHK | 82.3 |
| | AASIK | 84.5 |
| | AAAHK | 88.4 |
| | AAHRK | 80.1 |
| | AMWAK | 81.6 |
| | YATAK | 93.4 |
| | CAIAK | 90.3 |
| | PAEAK | 88.8 |
| | AAFRK | 81.6 |
| | YAFAK | 90.6 |
| | AAFWK | 82.3 |
| | AALVK | 85.4 |
| | YACAK | 81.8 |
| | ATFAK | 89.4 |
| | AAFKK | 90.9 |
| | AASRK | 83.0 |
| | EAVAR | 81.5 |
| | EATAR | 83.5 |
| | AALRR | 80.3 |
| | AAVRR | 81.9 |
| | TAFAR | 81.1 |
| | AAATK | 84.0 |
| | KAMAK | 86.9 |
| | KAIAK | 94.7 |
| | IATAK | 88.5 |
| **Cluster 2** | | |
| | AWMAD | 89.6 |
| | ASMAD | 81.7 |
| | AVLAD | 94.6 |
| | TAMAD | 93.8 |
| | CANAD | 88.0 |
| | ATIAD | 90.2 |
| | AIYAD | 91.6 |
| | ALLAD | 82.6 |
| | FAAAD | 92.7 |
| | HASAD | 92.8 |
| | AYHAD | 86.3 |
| | CAFAD | 86.8 |
| | ALFAD | 88.1 |
| | SASAD | 89.4 |
| | AMEAD | 80.1 |
| | AADHD | 88.6 |
| | AQIAD | 85.8 |
| | ALYAD | 84.3 |
| | AVMAD | 89.1 |
| | YAAAD | 86.9 |
| | YANAD | 94.6 |
| | IAIAD | 91.8 |
| | NACAD | 86.6 |
| | TAIAD | 83.3 |
| | HAVAD | 90.7 |
| | FANAD | 92.6 |
| | HAMAD | 91.1 |
| | WANAD | 92.8 |
| | ATFAD | 82.0 |
| | VATAD | 81.9 |
| | AMFAD | 94.3 |
| | AMMAD | 82.4 |
| | TAAAD | 96.4 |
| | AAYLD | 91.6 |
| | QAFAD | 87.4 |
| | ASFAD | 90.6 |
| | AQVAD | 81.8 |
| | FAHAD | 82.6 |
| | AMSAD | 92.0 |
| | AYLAD | 88.0 |
| | ASSAD | 91.7 |
| | HANAD | 84.4 |
| | IALAD | 94.7 |
| | CATAD | 85.4 |
| | TASAD | 93.6 |
| | TAVAD | 94.0 |
| | TAWAD | 84.6 |
| | FAFAD | 81.9 |
| | AAFLD | 82.9 |
| | MANAD | 95.3 |
| | IADAD | 92.3 |
| | MASAD | 91.5 |
| | FATAD | 91.9 |
| | LATAD | 91.9 |
| | AHTAD | 83.8 |
| | ARMAD | 86.3 |
| | YATAD | 85.2 |
| | AEWAD | 93.1 |
| | ACVAD | 95.4 |
| | AWWAD | 92.7 |
| | VAYAD | 82.9 |
| | VACAD | 84.8 |
| | SAIAD | 91.9 |
| | AHQAD | 86.5 |
| | MACAD | 95.2 |
| | AAVCD | 86.9 |
| | AWEAD | 93.4 |
| | AWFAD | 81.9 |
| | AFIAD | 87.4 |
| | FAYAD | 86.5 |
| | NATAD | 93.0 |
| | YAYAE | 80.4 |
| | FAQAD | 92.0 |
| | YAHAE | 84.3 |
| | WAVAE | 93.9 |
| | MAHAD | 85.5 |
| | WAVAD | 90.1 |
| | MAQAD | 83.5 |
| | TAFAD | 95.5 |
| | YAQAD | 81.8 |
| | MAFAD | 84.0 |
| | WAMAD | 92.1 |
| | YAIAD | 90.6 |
| | WAYAD | 92.8 |
| | WAIAD | 95.1 |
| | WAAAD | 91.9 |
| | NAVAD | 88.2 |
| | IAWAD | 89.2 |
| | AADCD | 80.1 |
| | AFDAD | 84.5 |
| | AADQD | 85.7 |
| | ASDAD | 85.9 |
| | AMTAD | 81.7 |
| | CAYAD | 83.1 |
| | AADND | 83.6 |
| | FADAD | 95.2 |
| | AAEQD | 81.2 |
| | AYDAD | 85.6 |
| | ALDAD | 82.3 |
| | SACAD | 80.3 |
| | AWCAD | 93.8 |
| | AADGD | 87.8 |
| | PAWAD | 84.8 |
| | AIVAD | 82.3 |
| | AATTD | 82.5 |
| | AISAD | 90.2 |
| | AALLD | 84.7 |
| | AFMAD | 92.4 |
| | AQLAD | 85.3 |
| | AATSD | 83.5 |
| | AWAAD | 85.9 |
| | ALMAD | 81.2 |
| | AFVAD | 81.9 |
| | AAVSD | 86.6 |
| | ALWAD | 83.2 |
| | AFCAD | 88.0 |
| | AYMAD | 96.7 |
| | ALIAD | 92.0 |
| | AYVAD | 81.1 |
| | AIMAD | 86.7 |
| | AQFAD | 87.0 |
| | IAAAD | 90.2 |
| | AASYD | 81.8 |
| | AAMCD | 81.6 |
| | SAWAD | 84.7 |
| | PAHAD | 84.6 |
| | AALTD | 85.9 |
| | AATYD | 87.0 |
| | AANAD | 85.5 |
| | AAQHD | 80.8 |
| | AAYMD | 83.0 |
| | AAIHD | 80.8 |
| | AAMVD | 81.5 |
| | AAFVD | 92.6 |
| | AAVTD | 81.2 |
| | AAYTD | 85.2 |
| | APSAD | 93.4 |
| | ANDAD | 80.9 |
| | AAFTD | 90.4 |
| | AWDAD | 88.1 |
| | AAVID | 81.3 |
| | AACHD | 85.2 |
| | RAFAE | 80.1 |
| | MADAD | 93.3 |
| | AADID | 85.2 |
| | QAMAD | 80.1 |
| | YAFAD | 82.6 |
| | YALAD | 86.7 |
| | NAYAD | 88.4 |
| | IASAD | 85.1 |
| | WATAD | 97.4 |
| | IACAD | 82.6 |
| | FAMAD | 85.8 |
| | AVFAD | 88.5 |
| | AWIAD | 92.6 |
| | IAQAD | 82.3 |
| | WAQAD | 82.5 |
| | WACAD | 97.8 |
| | QALAD | 84.6 |
| | WAWAD | 98.0 |
| | FASAD | 93.6 |
| | LAMAD | 80.3 |
| | FAIAD | 81.4 |
| | MAFAE | 84.9 |
| | TAFAE | 89.2 |
| | MAHAE | 91.0 |
| | MANAE | 91.9 |
| | NAWAE | 88.2 |
| | AMIAE | 81.6 |
| | YALAE | 82.2 |
| | MALAE | 88.1 |
| | WATAE | 95.6 |
| | NANAE | 94.6 |
| | YACAE | 89.7 |
| | HAHAE | 91.9 |
| | WAHAE | 95.3 |
| | HADAE | 87.7 |
| | AYCAE | 81.5 |
| | MAIAE | 89.5 |
| | NANAD | 91.5 |
| | WADAD | 93.1 |
| | WAYAE | 86.2 |
| | WAWAE | 97.5 |
| | VAMAE | 80.5 |
| | WADAE | 85.6 |
| | WASAE | 94.4 |
| | WAMAE | 80.4 |
| | WAIAE | 92.4 |
| | IAVAE | 85.1 |
| | QAVAE | 81.6 |
| | MAVAE | 88.7 |
| | QAHAE | 84.0 |
| | TADAE | 86.8 |
| | VALAD | 82.1 |
| | YAVAD | 90.9 |
| | YAWAD | 87.4 |
| | YAMAD | 93.9 |
| | WASAD | 82.7 |
| | YAHAD | 81.1 |
| | HAMAE | 82.7 |
| | DANAE | 94.2 |
| | RAVAE | 87.5 |
| | RAYAE | 94.6 |
| | TATAE | 84.4 |
| | IAQAE | 91.3 |
| | NAMAE | 85.6 |
| | RAMAE | 87.0 |
| | RAIAE | 87.3 |
| | AWVAE | 81.5 |
| | RAWAE | 89.4 |
| | APSAE | 84.8 |
| | APYAE | 97.4 |
| | APNAE | 96.9 |
| | APLAD | 86.7 |
| | ARIAD | 81.8 |
| | AADFD | 91.0 |
| **Cluster 3** | | |
| | DALAS | 85.5 |
| | DASAF | 81.8 |
| | DAWAN | 90.7 |
| | DAFAC | 90.6 |
| | DWWAA | 83.1 |
| | DRQAA | 90.7 |
| | DATAT | 80.3 |
| | DAWAT | 89.5 |
| | DAWAL | 82.3 |
| | DAEAT | 94.6 |
| | DAEWA | 95.1 |
| | DTEAA | 97.0 |
| | DAEAL | 82.8 |
| | EAWAA | 80.3 |
| | DAWAH | 89.9 |
| | EAWAC | 87.6 |
| | DACAQ | 82.2 |
| | DALAN | 81.6 |
| | DADAL | 82.7 |
| | DAVAC | 86.8 |
| | DADAT | 80.4 |
| | DALAC | 81.8 |
| | DAMAC | 81.6 |
| | DNHAA | 80.5 |
| | DAIAS | 81.9 |
| | DAWAM | 86.1 |
| | DALAH | 83.0 |
| | DYWAA | 86.1 |
| | DAYAF | 80.8 |
| | DCWAA | 82.2 |
| | DSYAA | 89.6 |
| | DAYWA | 80.4 |
| | DASAT | 87.1 |
| | DAIAC | 81.1 |
| | DSEAA | 93.3 |
| | DAEAH | 94.4 |
| | DCEAA | 93.0 |
| | DLEAA | 90.8 |
| | DVEAA | 82.8 |
| | DYEAA | 93.4 |
| | EMEAA | 82.9 |
| | DWEAA | 87.8 |
| | DIEAA | 88.1 |
| | DATAM | 80.7 |
| | DACAH | 84.6 |
| | DAIAT | 88.6 |
| | DATAC | 82.8 |
| | DAEAQ | 96.4 |
| | DAYAT | 83.3 |
| | DAAAH | 85.8 |
| | DASAH | 89.6 |
| | DACAC | 89.9 |
| | DQEAA | 91.9 |
| | DAEAM | 95.1 |
| | DFEAA | 93.5 |
| | DAAAQ | 84.0 |
| | DAMAH | 95.2 |
| | DHCAA | 81.3 |
| | DAAAL | 85.6 |
| | DANAH | 94.1 |
| | DAFAH | 82.2 |
| | DAHAH | 85.9 |
| | DHHAA | 80.7 |
| | DAYAS | 80.3 |
| | DALAY | 80.7 |
| | DHNAA | 81.7 |
| | DEWAA | 87.3 |
| | DAMAN | 81.8 |
| | DEYAA | 82.8 |
| | DEDAA | 86.5 |
| | DEIAA | 83.3 |
| | DAVAS | 81.1 |
| | DAQAS | 83.4 |
| | DQWAA | 81.5 |
| | DAWYA | 83.3 |
| | DHWAA | 88.4 |
| | DAWAQ | 80.7 |
| | EKGAA | 86.7 |
| | DAMFA | 81.5 |
| | DAAAS | 82.2 |
| | DKTAA | 94.1 |
| | DKFAA | 97.3 |
| | DAFVA | 80.7 |
| | DAYVA | 86.4 |
| | EAWAS | 86.8 |
| | DHLAA | 87.0 |
| | DAWAC | 96.8 |
| | EKLAA | 89.5 |
| | ERHAA | 93.7 |
| | DAFAT | 88.3 |
| | EAEAH | 85.2 |
| | EAEAM | 96.0 |
| | DAEVA | 81.8 |
| | DAEAI | 88.9 |
| | DAEYA | 94.0 |
| | EASAK | 82.3 |
| | DAENA | 88.2 |
| | DAELA | 82.2 |
| | DAECA | 86.4 |
| | DAEIA | 89.9 |
| | EKQAA | 85.8 |
| | EAEAY | 88.3 |
| | DAEAV | 84.5 |
| | DAEQA | 88.6 |
| | DAEAA | 86.4 |
| | DAEMA | 94.9 |
| | DMEAA | 90.0 |
| | DAEAC | 98.7 |
| | EAEAV | 80.8 |
| | EAEAS | 98.3 |
| | EAEQA | 86.4 |
| | EAEAA | 87.3 |
| | EAENA | 84.4 |
| | EAEHA | 88.8 |
| | ECEAA | 88.0 |
| | EFEAA | 92.8 |
| | EAEAD | 80.7 |
| | EIEAA | 89.2 |
| | EAEVA | 86.8 |
| | EAEIA | 87.5 |
| | EAWAT | 86.8 |
| | EAEYA | 83.2 |
| | ECWAA | 80.1 |
| | DAEAF | 96.1 |
| | EAEAI | 82.3 |
| | EAEAT | 87.8 |
| | EAEAF | 80.5 |
| | ERQAA | 88.0 |
| | ESEAA | 88.2 |
| | EYEAA | 91.1 |
| | EAWHA | 85.0 |
| | EKFAA | 91.6 |
| | EAWIA | 82.5 |
| | EAWAM | 86.8 |
| | EAWSA | 80.6 |
| | DAEAN | 93.3 |
| | DKKAA | 82.5 |
| | DRFAA | 87.2 |
| | DRSAA | 89.0 |
| | DRAAA | 93.0 |
| | DRTAA | 96.4 |
| | DKVAA | 92.4 |
| | DKCAA | 91.9 |
| | DKWAA | 98.7 |
| | DKYAA | 94.6 |
| | DAMAF | 85.3 |
| | DKQAA | 92.0 |
| | DAWAS | 94.8 |
| | DKLAA | 97.5 |
| | DADAC | 92.1 |
| | DAESA | 93.3 |
| | DAEAS | 98.8 |
| | DAYAC | 90.9 |
| | DFWAA | 83.9 |
| | DAEFA | 91.8 |
| | DAEHA | 96.7 |
| | DKHAA | 97.4 |
| | DAAAC | 83.8 |
| | DALAT | 90.5 |
| | DRIAA | 91.0 |
| | DRNAA | 96.8 |
| | DRGAA | 81.0 |
| | DRYAA | 93.7 |
| | DAWIA | 82.6 |
| | DRVAA | 99.3 |
| | DRMAA | 87.5 |
| | DKMAA | 96.0 |
| | DKAAA | 97.5 |
| | EKSAA | 93.2 |
| | DKSAA | 96.5 |
| | EEQAA | 84.4 |
| | EAESA | 92.2 |
| | EAEAQ | 93.4 |
| | HAEAE | 90.8 |
| | EETAA | 81.1 |
| | DEVAA | 91.1 |
| | KADAE | 94.8 |
| | KAEAE | 95.3 |
| | RAEAE | 80.0 |
| | EATAK | 93.5 |
| | EKNAA | 85.4 |
| | EKAAA | 93.9 |
| | EKCAA | 86.6 |
| | EKHAA | 88.9 |
| | DKRAA | 86.5 |
| | KRDAA | 88.9 |
| **Cluster 4** | | |
| | RAHAK | 96.6 |
| | RLLAA | 85.8 |
| | RATAK | 89.8 |
| | KAFAK | 89.8 |
| | KATAK | 90.6 |
| | RAYAK | 96.1 |
| | RWWAA | 85.7 |
| | RRCAA | 90.0 |
| | RAWAK | 97.0 |
| | RKMAA | 90.4 |
| | RRIAA | 90.4 |
| | KAYAK | 93.8 |
| | RASAR | 80.8 |
| | RKFAA | 94.3 |
| | RLIAA | 86.4 |
| | RASAK | 84.0 |
| | RAQAK | 91.5 |
| | RALAR | 95.3 |
| | KAWAK | 93.6 |
| | KKWAA | 81.4 |
| | KRFAA | 95.7 |
| | KAVAK | 96.0 |
| | RKCAA | 86.0 |
| | RKSAA | 93.0 |
| | PAIRA | 85.1 |
| | RYIAA | 86.8 |
| | RFFAA | 88.3 |
| | RRSAA | 90.7 |
| | RRFAA | 87.6 |
| | RRAAA | 97.1 |
| | RIVAA | 81.8 |
| | RAFAR | 99.0 |
| | RYFAA | 86.1 |
| | RKNAA | 84.7 |
| | RAVAK | 92.8 |
| | RRRAA | 89.2 |
| | RARAK | 90.6 |
| | RRYAA | 85.5 |
| | RKIAA | 87.9 |
| | RKTAA | 90.6 |
| | RVVAA | 80.2 |
| | RAADA | 88.3 |
| | RATSA | 83.9 |
| | RLFAA | 84.6 |
| | KAIAR | 86.2 |
| | KARAR | 90.9 |
| | RKLAA | 95.3 |
| | KKLAA | 97.0 |
| | RKAAA | 95.4 |
| | RRVAA | 98.8 |
| | KALAK | 84.6 |
| | KKTAA | 88.8 |
| | KAHAK | 93.8 |
| | KKCAA | 91.3 |
| | KKFAA | 91.7 |
| | RAHAR | 83.7 |
| | RVMAA | 85.0 |
| | RAKEA | 85.8 |
| | RAWAR | 89.3 |
| | REFAA | 87.9 |
| | TKIAA | 90.0 |
| | PKEAA | 93.0 |
| | PAWRA | 82.6 |
| | ANEDA | 80.9 |
| | EALDA | 84.3 |
| | EAVDA | 95.1 |
| | EACEA | 81.4 |
| | EAYEA | 85.1 |
| | EATEA | 89.7 |
| | MEFAA | 81.4 |
| | TEFAA | 84.7 |
| | EAMEA | 83.2 |
| | YAKEA | 91.0 |
| | EARIA | 94.5 |
| | NACEA | 80.7 |
| | WELAA | 88.2 |
| | WENAA | 88.8 |
| | HEFAA | 81.1 |
| | MEMAA | 82.2 |
| | EAAEA | 91.0 |
| | DARCA | 82.0 |
| | DAFEA | 96.3 |
| | KAKAE | 92.1 |
| | NKEAA | 80.7 |
| | WNDAA | 89.4 |
| | WKEAA | 82.4 |
| | WHEAA | 92.7 |
| | WNEAA | 94.0 |
| | WADAA | 86.0 |
| | WCDAA | 80.3 |
| | WIDAA | 89.1 |
| | RAEAK | 93.6 |
| **Cluster 5** | | |
| | KAFAD | 89.5 |
| | RATAD | 93.0 |
| | NAFAD | 86.7 |
| | KAWAE | 82.5 |
| | QATAD | 97.0 |
| | RAYAD | 93.5 |
| | RAQAD | 95.2 |
| | RAIAD | 96.2 |
| | EAYAD | 83.3 |
| | MAYAD | 95.2 |
| | QAYAD | 83.1 |
| | KAAAD | 85.4 |
| | NAIAD | 88.5 |
| | SATAD | 89.8 |
| | KAVAD | 98.8 |
| | HAFAD | 96.0 |
| | VAWAD | 81.6 |
| | QASAD | 95.0 |
| | HAQAD | 92.0 |
| | IAFAE | 86.4 |
| | RAWAD | 96.9 |
| | KATAE | 88.6 |
| | RAHAD | 96.5 |
| | KASAD | 91.0 |
| | RALAD | 95.1 |
| | RASAD | 96.1 |
| | MALAD | 97.6 |
| | KACAD | 92.0 |
| | RAFAD | 87.8 |
| | KAMAD | 98.2 |
| | RANAD | 81.2 |
| | NAAAD | 90.2 |
| | NASAD | 96.2 |
| | NADAD | 91.3 |
| | NAMAD | 89.5 |
| | LAYAD | 95.6 |
| | FASAE | 80.2 |
| | VAFAE | 89.0 |
| | VALAE | 82.7 |
| | AANNE | 82.9 |
| | SATAE | 85.3 |
| | KAHAD | 90.3 |
| | QAFAE | 97.8 |
| | KASAE | 84.2 |
| | QAIAE | 94.4 |
| | NAIAE | 98.8 |
| | IASAE | 97.4 |
| | LALAE | 98.8 |
| | AAHRE | 91.0 |
| | IACAE | 98.2 |
| | IAWAE | 98.0 |
| | TAHAE | 95.3 |
| | HALAE | 95.3 |
| | NAYAE | 91.1 |
| | TAWAE | 96.2 |
| | CALAE | 87.1 |
| | AAMRE | 90.0 |
| | ASYAE | 87.2 |
| | IATAE | 99.6 |
| | FAMAE | 87.1 |
| | NACAE | 85.7 |
| | TANAE | 96.6 |
| | EAYAE | 98.9 |
| | QAQAE | 89.0 |
| | TACAE | 96.4 |
| | NAWAD | 96.5 |
| | KALAD | 84.2 |
| | TATAD | 97.9 |
| | NATAE | 90.4 |
| | EAIAE | 93.0 |
| | VANAE | 93.7 |
| | HAIAD | 88.5 |
| | RAVAD | 92.5 |
| | NAHAD | 83.8 |
| | HAAAD | 87.3 |
| | TASAE | 88.1 |
| | HADAD | 92.7 |
| | AAFRE | 80.4 |
| | AAQRE | 91.7 |
| | AACKE | 88.3 |
| | AASRE | 88.0 |
| | AAYRE | 91.7 |
| | IAIAE | 84.2 |
| | TAIAE | 91.9 |
| | AAMKE | 92.3 |
| | AARRE | 95.8 |
| **Cluster 6** | | |
| | KEHAA | 85.6 |
| | KECAA | 85.9 |
| | KESAA | 90.3 |
| | KELAA | 87.7 |
| | KEWAA | 86.3 |
| | REQAA | 90.4 |
| | REMAA | 91.6 |
| | REWAA | 95.2 |
| | RECAA | 95.6 |
| | KEMAA | 93.0 |
| | REVAA | 89.0 |
| | REHAA | 91.3 |
| | REIAA | 90.5 |
| | RESAA | 96.6 |
| | KENAA | 81.9 |
| | KAFEA | 81.6 |
| | KAMEA | 87.9 |
| | KEQAA | 86.2 |
| | KEFAA | 91.9 |
| | KEIAA | 91.0 |
| | KAREA | 95.1 |
| | RENAA | 96.3 |
| | REAAA | 98.2 |
| | REYAA | 90.6 |
| | RETAA | 81.3 |
| | RDAAA | 91.6 |
| | RDWAA | 97.7 |
| | RDMAA | 96.6 |
| | RDCAA | 92.6 |
| | RDVAA | 98.6 |
| | KEVAA | 88.8 |
| | KAIEA | 93.7 |
| | RDFAA | 82.8 |
| | RDSAA | 92.8 |
| | RDIAA | 90.9 |
| | RDTAA | 97.9 |
| | KAQEA | 87.7 |
| | KDIAA | 87.3 |
| | KAECA | 84.3 |
| | REGAA | 84.5 |
| | KEGAA | 84.8 |
| **Cluster 7** | | |
| | KRYAA | 86.1 |
| | KYYAA | 81.4 |
| | KALWA | 98.3 |
| | KAYIA | 93.4 |
| | KAIAA | 80.1 |
| | KAFAI | 94.6 |
| | KAGAH | 82.9 |
| | KAGLA | 86.0 |
| | KAGAY | 83.5 |
| | KFGAA | 87.1 |
| | KHGAA | 81.6 |
| | KLYAA | 93.0 |
| | KWVAA | 96.5 |
| | KFCAA | 83.4 |
| | KAIVA | 96.5 |
| | KMGAA | 83.9 |
| | KFYAA | 96.3 |
| | KLGAA | 81.5 |
| | KWGAA | 84.0 |
| | RFYAA | 93.1 |
| | KTLAA | 94.2 |
| | RQVAA | 89.3 |
| | KAIAW | 96.7 |
| | KAGAL | 89.7 |
| | RMIAA | 86.1 |
| | KSFAA | 91.6 |
| | RAWYA | 88.8 |
| | RAWIA | 93.3 |
| | RALAC | 90.1 |
| | RAILA | 82.7 |
| | RALAV | 87.3 |
| | KAGAI | 84.0 |
| | RRGAA | 80.1 |
| | RKGAA | 87.6 |
| | RAGIA | 82.8 |
| | RALAY | 87.0 |
| | RAEEA | 90.2 |
| | NAIEA | 94.9 |
| **Cluster 8** | | |
| | FAPAK | 81.2 |
| | AAPIK | 90.8 |
| | AAPSK | 89.7 |
| | VAPAK | 90.4 |
| | AAPCK | 83.4 |
| | LAPAK | 80.1 |
| | ASPAK | 81.8 |
| | AAPRK | 87.7 |
| | AAPTK | 91.3 |
| | AAPQK | 80.6 |
| | ARPAK | 87.3 |
| | AHPAK | 81.8 |
| | AAPYK | 88.0 |
| | AAPMK | 81.1 |
| | AAPHK | 87.4 |
| | AVPAK | 80.8 |
| | AAPKK | 86.7 |
| | ARRAK | 88.6 |
| | AFPAR | 83.7 |
| | TAPAK | 87.3 |
| | AAPQR | 80.1 |
| | CAPAK | 80.8 |
| | AAPLK | 86.2 |
| | AYPAK | 92.4 |
| | AKQAK | 80.1 |
| | AKWAR | 87.8 |
| | AKTAK | 80.2 |
| | AKPAK | 83.0 |
| | AKVAK | 90.2 |
| **Cluster 9** | | |
| | DANAP | 94.9 |
| | DAAAP | 92.8 |
| | DATAP | 95.5 |
| | DAVAP | 92.5 |
| | DAHAP | 94.5 |
| | DACAP | 91.0 |
| | DAWAP | 82.5 |
| | EAYAP | 92.2 |
| | DARAP | 89.7 |
| | EALAP | 90.4 |
| | DAYAP | 95.8 |
| | DAFAP | 91.3 |
| | DALAP | 95.6 |
| | EASAP | 92.4 |
| | DAQAP | 98.1 |
| | EAFAP | 91.5 |
| | EAIAP | 88.2 |
| | EAMAP | 93.1 |
| | EAAAP | 80.7 |
| | DASAP | 89.9 |
| | DAMAP | 93.7 |
| | EAHAP | 85.5 |
| | EAVAP | 93.0 |
| | EANAP | 94.6 |
| | APEAP | 84.7 |
| | EWIAA | 80.1 |
| | APYRA | 83.1 |
| | EARAR | 86.5 |
| **Cluster 10** | | |
| | WDPAA | 92.8 |
| | RAPMA | 83.3 |
| | KAPMA | 83.6 |
| | APPEA | 80.8 |
| | RAPIA | 80.7 |
| | RAPWA | 95.0 |
| | RNPAA | 87.6 |
| | RAPAS | 84.3 |
| | RAPAE | 93.2 |
| | KNPAA | 82.7 |
| | KAPTA | 85.1 |
| | RAPPA | 92.5 |
| | RAPYA | 86.3 |
| | KAPPA | 95.7 |
| | DDRAA | 91.5 |
| | EDKAA | 92.1 |
| | KDDAA | 87.5 |
| | KAYPA | 97.2 |
| | KEPAA | 83.9 |
| | KAQAR | 86.2 |
| | KAPAD | 92.1 |
| | KWEAA | 88.0 |
| | KVEAA | 91.2 |
| | KDTAA | 84.5 |
| | KEYAA | 99.0 |
| | KDMAA | 94.6 |
| | KDYAA | 97.5 |
| | KDFAA | 82.3 |
| **Cluster 11** | | |
| | DAIAR | 84.1 |
| | DAYAR | 90.0 |
| | DAHAK | 96.8 |
| | DAQAR | 80.9 |
| | DACAK | 92.6 |
| | DATAR | 89.2 |
| | DAVAR | 93.8 |
| | DANAR | 81.3 |
| | EAWAR | 97.6 |
| | DALAR | 86.4 |
| | DAFAK | 90.5 |
| | DASAR | 85.7 |
| | DATAK | 89.0 |
| | DAIAK | 93.0 |
| | DAWAR | 89.7 |
| | DVCAA | 83.4 |
| | DDFAA | 92.9 |
| | EACAK | 89.0 |
| **Cluster 12** | | |
| | KAKRA | 88.8 |
| | KIKAA | 81.8 |
| | KAKYA | 91.6 |
| | KPTAA | 90.9 |
| | KATRA | 82.4 |
| | KAKKA | 99.0 |
| | RPKAA | 86.7 |
| | KPQAA | 90.5 |
| | KPYAA | 90.7 |
| | KACRA | 89.8 |
| | RPHAA | 84.3 |
| | KPFAA | 92.7 |
| | RAKRA | 88.7 |
| | RAHRA | 89.8 |
| | EPPAA | 93.2 |
| | KPNAA | 91.1 |
| | KALRA | 81.8 |
| **Cluster 13** | | |
| | DAKAI | 86.6 |
| | ESQAA | 97.2 |
| | DAKAN | 91.3 |
| | DAKAC | 94.0 |
| | DAKAM | 82.3 |
| | ECYAA | 94.8 |
| | ECFAA | 95.8 |
| | DAQNA | 85.1 |
| | DFIAA | 95.7 |
| | DLFAA | 99.3 |
| | DCMAA | 96.5 |
| | DGYAA | 83.0 |
| | DGRAA | 88.6 |
| | DGKAA | 88.0 |
| | DGHAA | 87.0 |
| **Cluster 14** | | |
| | AARKS | 81.3 |
| | AARKI | 81.3 |
| | VARKA | 87.7 |
| | AIRKA | 80.4 |
| | AWRKA | 86.5 |
| | AAKKF | 80.7 |
| | AAKRV | 82.5 |
| | ADRWA | 83.6 |
| | AERFA | 82.9 |
| | IDRAA | 83.1 |
| | AAKRR | 82.4 |
| | EARPA | 83.3 |
| | RDRAA | 88.3 |
| **Cluster 15** | | |
| | APWAD | 93.8 |
| | APQAD | 95.0 |
| | APYAD | 91.5 |
| | APPAD | 80.7 |
| | APRAE | 91.6 |
| | APKAE | 96.8 |
| | APWAE | 92.3 |
| | APIAE | 96.5 |
| | APEAK | 94.7 |
| | APPAE | 91.5 |
| **Cluster 16** | | |
| | ADRQA | 84.6 |
| | ADKHA | 81.1 |
| | ADKAT | 84.2 |
| | ADRRA | 87.0 |
| | VEKAA | 82.8 |
| | ADRAL | 83.2 |
| | ADRTA | 82.7 |
| | RKRAA | 84.0 |
| | KKRAA | 94.5 |
| | RRKAA | 90.7 |
| **Cluster 17** | | |
| | PADTA | 85.8 |
| | PYDAA | 83.3 |
| | PADWA | 86.6 |
| | PADAW | 83.4 |
| | PADAM | 83.3 |
| | PADAY | 85.3 |
| | PADEA | 89.1 |
| | PRDAA | 82.4 |
| | PAEEA | 84.6 |
| **Cluster 18** | | |
| | KANAR | 82.0 |
| | KAAAR | 80.5 |
| | KAFAR | 85.8 |
| | KAVRA | 80.2 |
| | KASRA | 89.8 |
| | ADQAP | 82.0 |
| | ACEAP | 81.4 |
| | AYEAP | 82.4 |
| | AHEAP | 81.5 |
| **Cluster 19** | | |
| | APGTA | 80.2 |
| | APGVA | 80.3 |
| | APGLA | 83.7 |
| | APGCA | 80.9 |
| | TPGAA | 81.3 |
| | APGMA | 85.6 |
| **Cluster 20** | | |
| | PDNAA | 82.0 |
| | PDTAA | 94.6 |
| | PDAAA | 82.1 |
| | PDIAA | 84.8 |
| **Cluster 21** | | |
| | EAHAW | 81.3 |
| | NAEAW | 85.4 |
| | EAIAK | 99.5 |
| **Cluster 22** | | |
| | RAEAW | 87.5 |
| | DAKKA | 90.5 |
| | KAWEA | 85.1 |
| **Cluster 23** | | |
| | NDIAA | 81.9 |
| | NDCAA | 84.6 |
| | HDFAA | 85.6 |
| **Cluster 24** | | |
| | APPAR | 88.0 |
| | APPAK | 80.5 |
| | DAPAK | 89.5 |
| **Cluster 25** | | |
| | AACDD | 80.8 |
| | AANDD | 81.5 |
| **Cluster 26** | | |
| | AEHRA | 84.9 |
| | AENKA | 86.8 |
| **Cluster 27** | | |
| | APEAE | 81.1 |
| | APDRA | 82.9 |
| **Cluster 28** | | |
| | DARGA | 81.6 |
| | DARQA | 93.9 |
| **Cluster 29** | | |
| | EKDAA | 84.0 |
| **Cluster 30** | | |
| | PAGAP | 80.8 |
| **Cluster 31** | | |
| | KNGAA | 84.6 |
| **Cluster 32** | | |
| | APPAP | 82.6 |
| **Cluster 33** | | |
| | KAKDA | 87.5 |
| **Cluster 34** | | |
| | EAREA | 86.1 |
| **Cluster 35** | | |
| | WAEYA | 81.0 |
| **Cluster 36** | | |
| | PAKKA | 82.6 |
| **Cluster 37** | | |
| | KGYAA | 87.7 |
| **Cluster 38** | | |
| | AAWCD | 81.9 |
| **Cluster 39** | | |
| | KDKAA | 90.8 |
| **Cluster 40** | | |
| | PAHVA | 83.9 |
| **Cluster 41** | | |
| | RARAY | 81.0 |
| **Cluster 42** | | |
| | AAPKE | 92.8 |
| **Cluster 43** | | |
| | APKGA | 81.3 |
| **Cluster 44** | | |
| | EAKAK | 95.4 |
| **Cluster 45** | | |
| | FDGAA | 84.0 |
| **Cluster 46** | | |
| | WAEAW | 81.3 |
| **Cluster 47** | | |
| | PDPAA | 83.7 |
| **Cluster 48** | | |
| | RAMAD | 94.5 |
| **Cluster 49** | | |
| | KREAA | 80.3 |
| **Cluster 50** | | |
| | PACAK | 80.1 |
| **Cluster 51** | | |
| | DAICA | 81.8 |
| **Cluster 52** | | |
| | RAEAD | 86.2 |
| **Cluster 53** | | |
| | APKKA | 82.3 |
| **Cluster 54** | | |
| | EAGKA | 86.6 |

**Supplementary Table 2.** 54 structural clusters formed by pentapeptides with stable spatial conformation. Cα-atoms of the first amino acid residue from the N-terminus is highlighted in green, and the Cα atoms of the first three N-terminal amino acid residues of pentapeptide are located in the vertical plane.

| Cluster number | The sequence of the representative pentapeptide structure | Number of elements in a cluster | Торсионные углы ϕ и ψ центр. ИЕ | | Representative spatial structure of the cluster. N-terminus is highlighted in green. The first three N-terminal Cα-atoms are located in the vertical plane. |
|---|---|---|---|---|---|
| 1 | AALFK | 324 | **ψ1** | -28.8 | 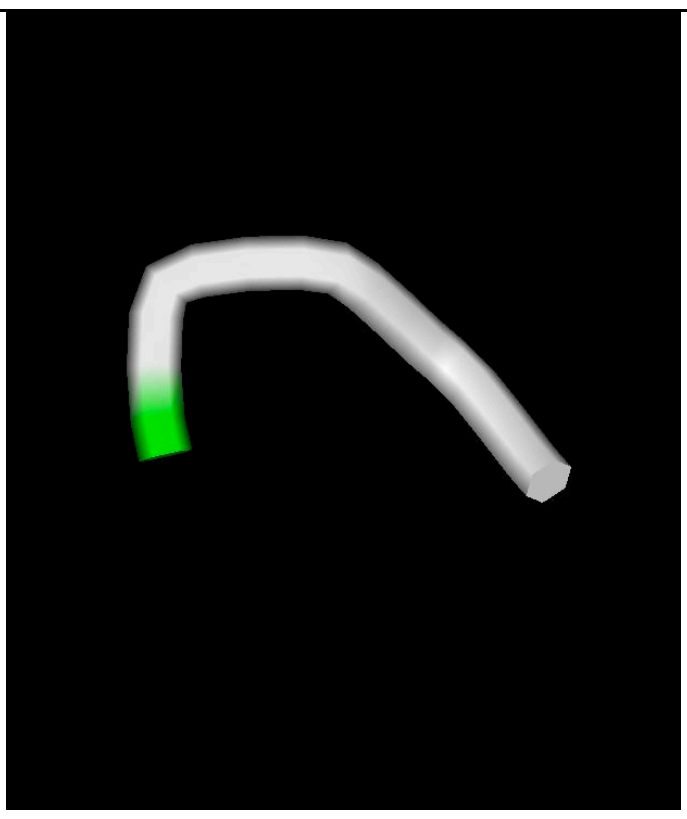 |
| | | | **φ2** | -81.9 | |
| | | | **ψ2** | -44.9 | |
| | | | **φ3** | -97.2 | |
| | | | **ψ3** | 156.1 | |
| | | | **φ4** | -80.4 | |
| | | | **ψ4** | 156.1 | |
| | | | **φ5** | -123. | |
| | | | | | |
| 2 | AWMAD | 216 | **ψ1** | 85.8 | 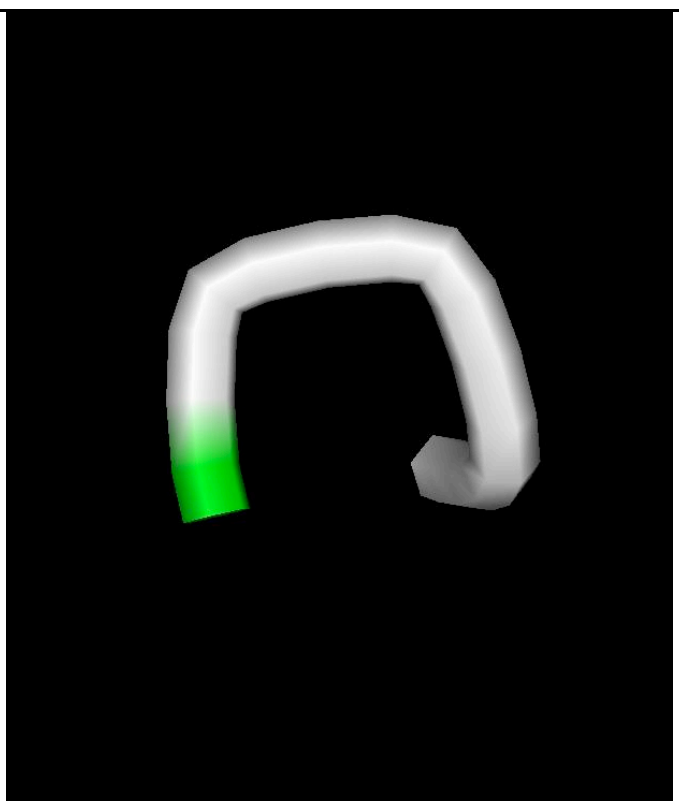 |
| | | | **φ2** | -99.3 | |
| | | | **ψ2** | -39.7 | |
| | | | **φ3** | -121.1 | |
| | | | **ψ3** | 44.4 | |
| | | | **φ4** | -155.3 | |
| | | | **ψ4** | 44.4 | |
| | | | **φ5** | -137.0 | |
| | | | | | |
| 3 | DALAS | 186 | **ψ1** | 135.8 | 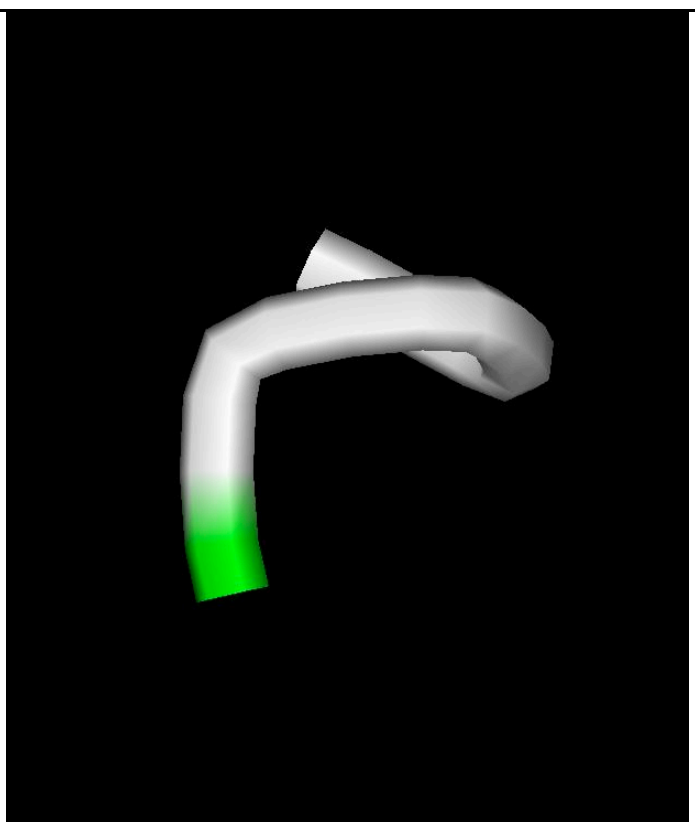 |
| | | | **φ2** | -138.3 | |
| | | | **ψ2** | 22.7 | |
| | | | **φ3** | -105.6 | |
| | | | **ψ3** | -14.1 | |
| | | | **φ4** | -110.6 | |
| | | | **ψ4** | -14.1 | |
| | | | **φ5** | -113.5 | |
| | | | | | |
| 4 | RAHAK | 92 | **ψ1** | 108.0 | 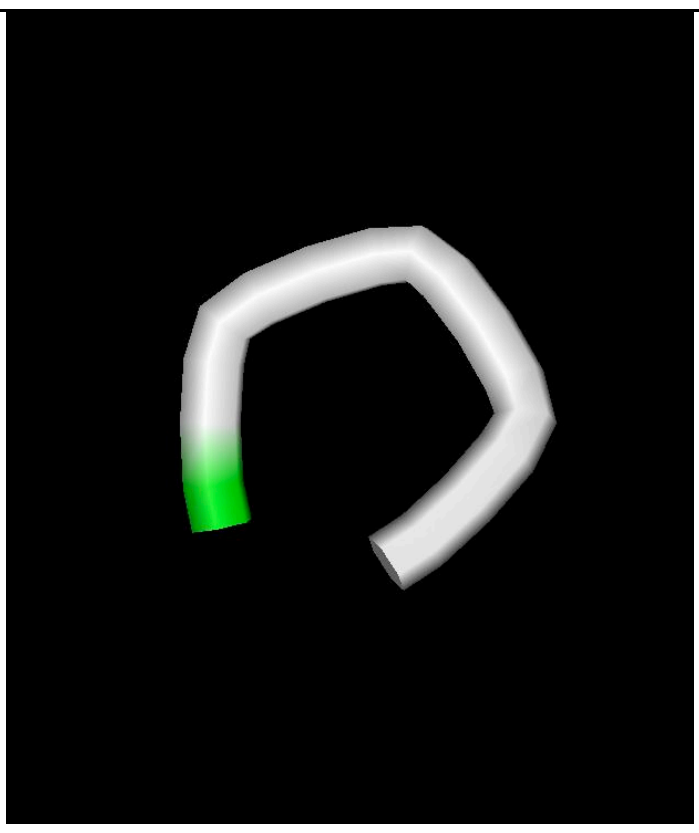 |
| | | | **φ2** | -117.4 | |
| | | | **ψ2** | -66.5 | |
| | | | **φ3** | -109.6 | |
| | | | **ψ3** | -51.1 | |
| | | | **φ4** | -110.9 | |
| | | | **ψ4** | -51.1 | |
| | | | **φ5** | -120.8 | |
| | | | | | |
| 5 | KAFAD | 86 | **ψ1** | -60.8 | 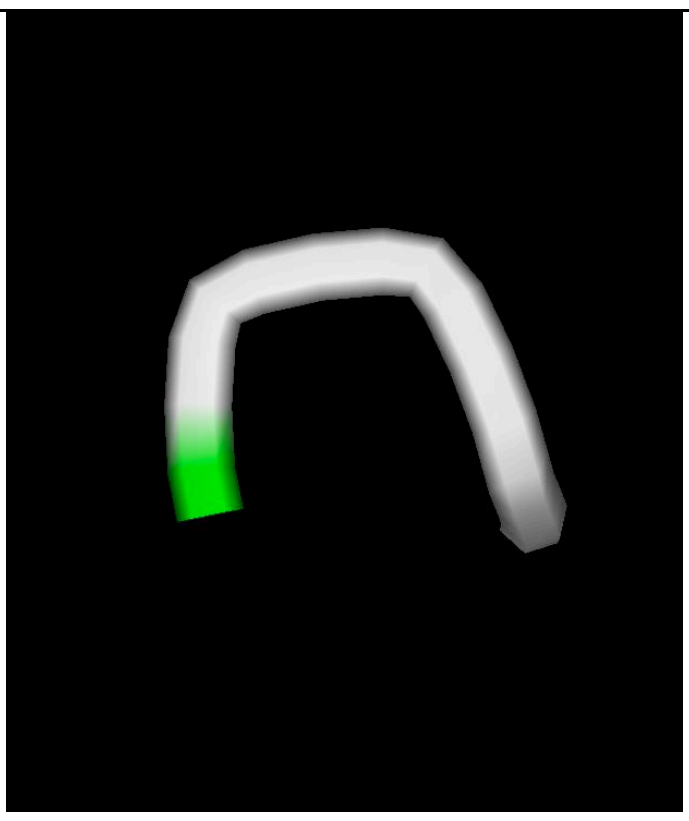 |
| | | | **φ2** | -99.9 | |
| | | | **ψ2** | -45.7 | |
| | | | **φ3** | -137.0 | |
| | | | **ψ3** | 61.5 | |
| | | | **φ4** | -159.7 | |
| | | | **ψ4** | 61.5 | |
| | | | **φ5** | -129.2 | |
| | | | | | |
| 6 | KEHAA | 41 | **ψ1** | -53.0 | 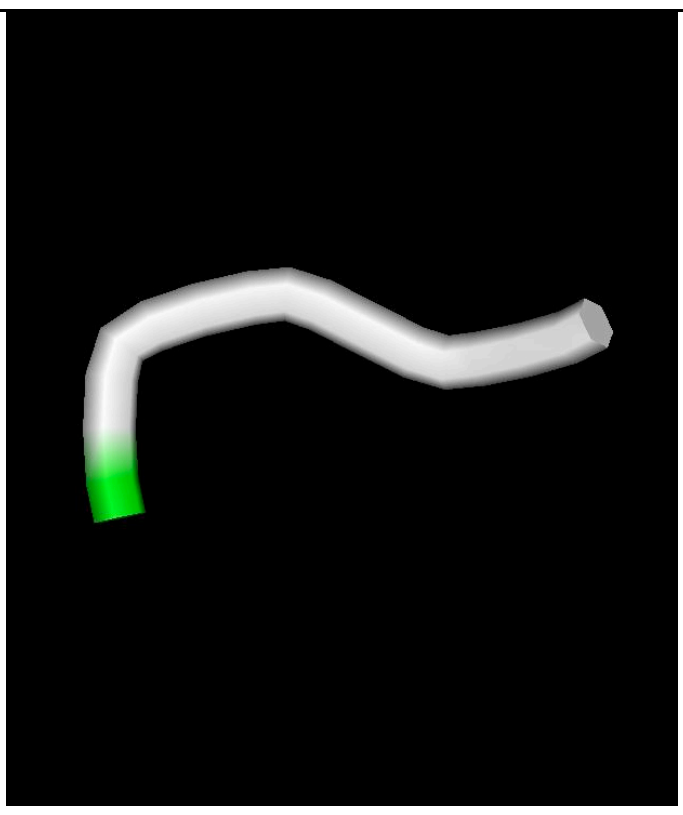 |
| | | | **φ2** | -124.6 | |
| | | | **ψ2** | -53.6 | |
| | | | **φ3** | -139.4 | |
| | | | **ψ3** | 169.6 | |
| | | | **φ4** | -122.2 | |
| | | | **ψ4** | 169.6 | |
| | | | **φ5** | -93.9 | |
| | | | | | |
| 7 | KRYAA | 38 | **ψ1** | 127.0 | 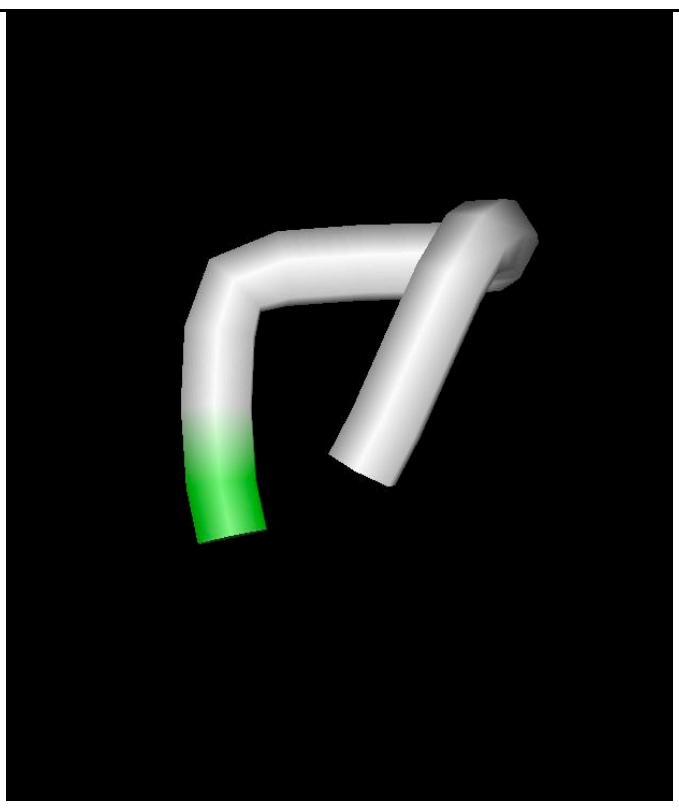 |
| | | | **φ2** | -113.0 | |
| | | | **ψ2** | 54.5 | |
| | | | **φ3** | 68.7 | |
| | | | **ψ3** | -38.3 | |
| | | | **φ4** | -87.5 | |
| | | | **ψ4** | -38.3 | |
| | | | **φ5** | -92.2 | |
| | | | | | |
| 8 | FAPAK | 29 | **ψ1** | -56.9 | 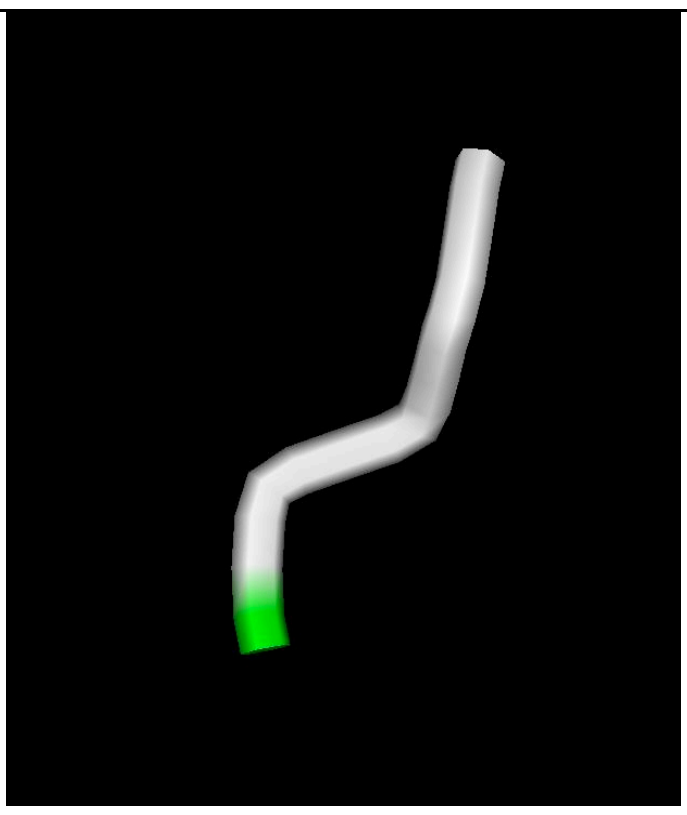 |
| | | | **φ2** | -89.8 | |
| | | | **ψ2** | 129.0 | |
| | | | **φ3** | -67.7 | |
| | | | **ψ3** | 137.9 | |
| | | | **φ4** | -79.8 | |
| | | | **ψ4** | 137.9 | |
| | | | **φ5** | -121.2 | |
| | | | | | |
| 9 | DANAP | 28 | **ψ1** | 166.3 | 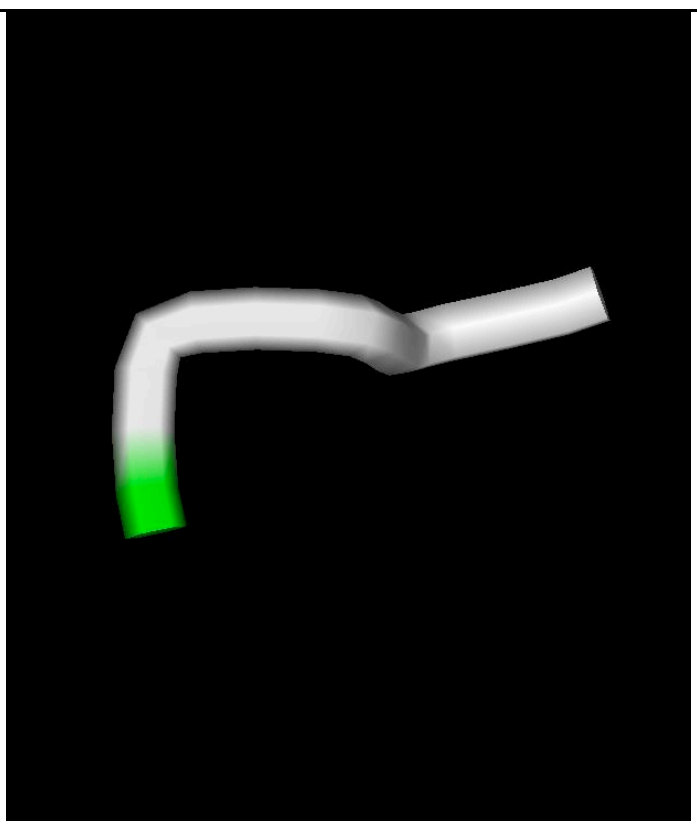 |
| | | | **φ2** | -106.4 | |
| | | | **ψ2** | -8.9 | |
| | | | **φ3** | -94.1 | |
| | | | **ψ3** | 100.5 | |
| | | | **φ4** | -119.4 | |
| | | | **ψ4** | 100.5 | |
| | | | **φ5** | -82.2 | |
| | | | | | |
| 10 | WDPAA | 28 | **ψ1** | 133.6 | 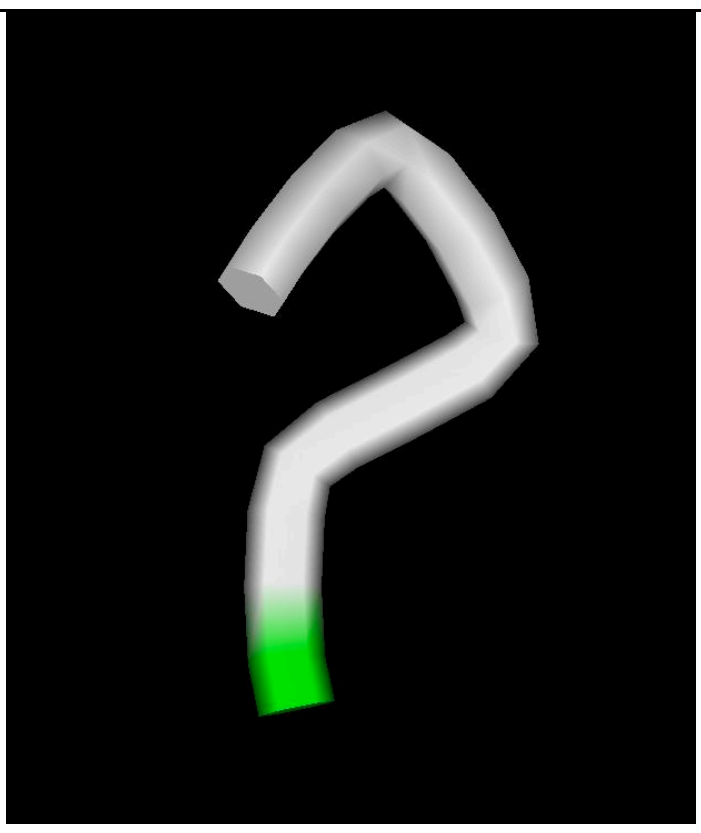 |
| | | | **φ2** | -101.5 | |
| | | | **ψ2** | 144.5 | |
| | | | **φ3** | -81.1 | |
| | | | **ψ3** | -20.4 | |
| | | | **φ4** | -104.4 | |
| | | | **ψ4** | -20.4 | |
| | | | **φ5** | -131.0 | |
| | | | | | |
| 11 | DAIAR | 18 | **ψ1** | 143.2 | 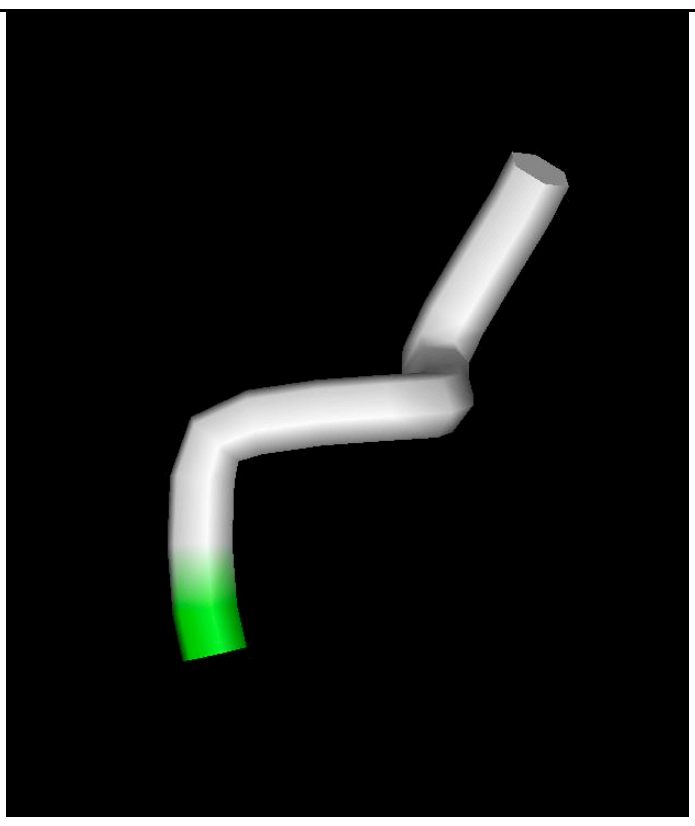 |
| | | | **φ2** | -135.8 | |
| | | | **ψ2** | 23.7 | |
| | | | **φ3** | -104.1 | |
| | | | **ψ3** | 53.9 | |
| | | | **φ4** | -106.2 | |
| | | | **ψ4** | 53.9 | |
| | | | **φ5** | -132.0 | |
| | | | | | |
| 12 | KAKRA | 17 | **ψ1** | 97.3 | 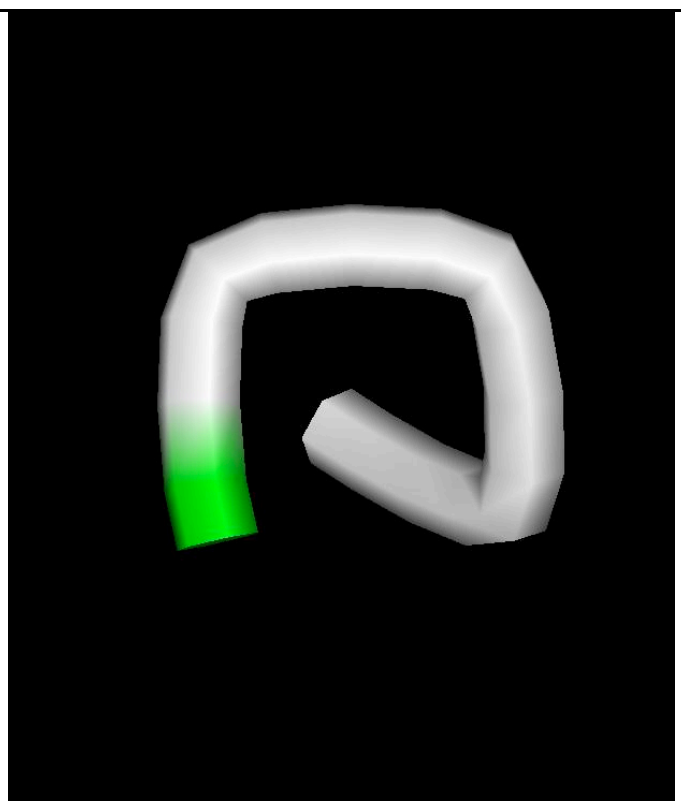 |
| | | | **φ2** | -72.9 | |
| | | | **ψ2** | -38.7 | |
| | | | **φ3** | -83.6 | |
| | | | **ψ3** | -39.2 | |
| | | | **φ4** | -67.1 | |
| | | | **ψ4** | -39.2 | |
| | | | **φ5** | -106.8 | |
| | | | | | |
| 13 | DAKAI | 15 | **ψ1** | -54.8 | 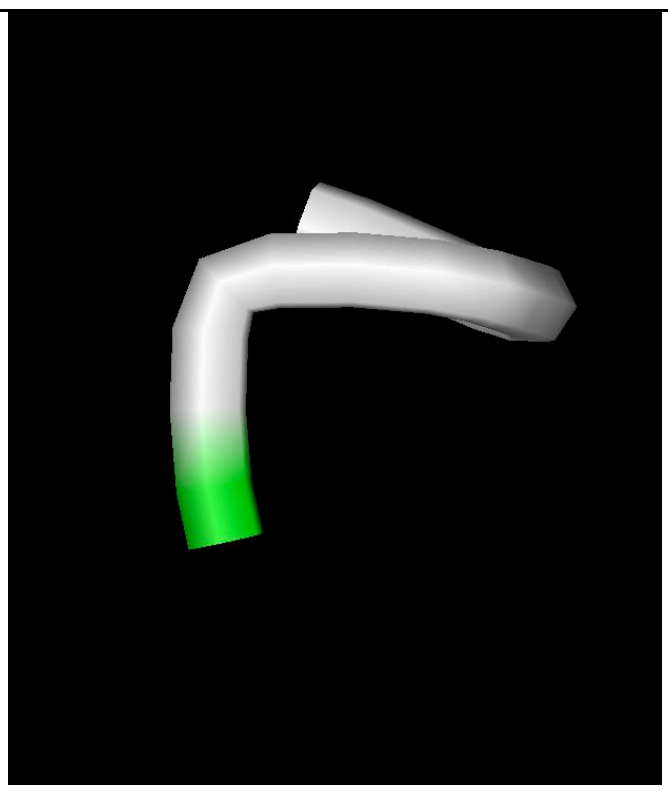 |
| | | | **φ2** | 71.6 | |
| | | | **ψ2** | 35.0 | |
| | | | **φ3** | -130.9 | |
| | | | **ψ3** | -31.7 | |
| | | | **φ4** | -109.0 | |
| | | | **ψ4** | -31.7 | |
| | | | **φ5** | -110.8 | |
| | | | | | |
| 14 | AARKS | 13 | **ψ1** | -37.5 | 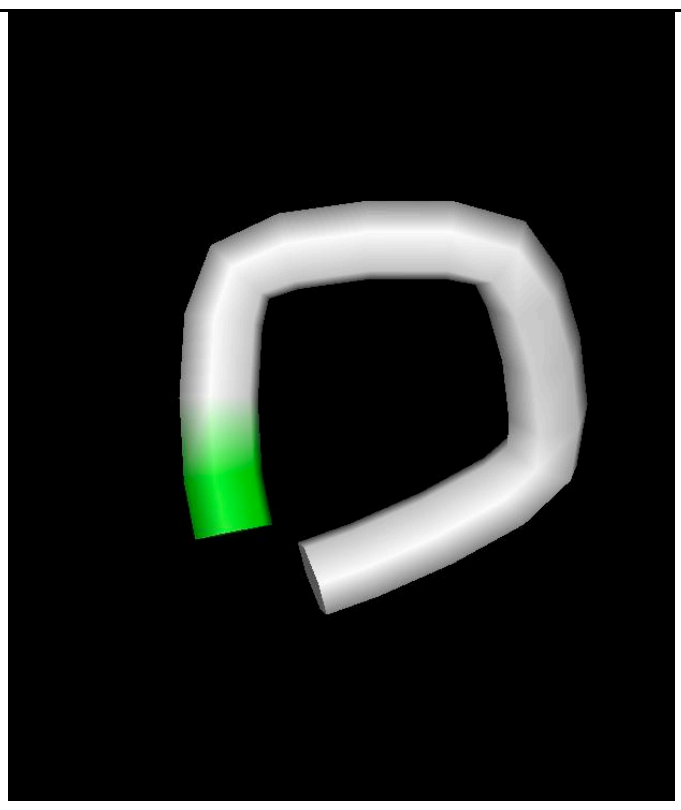 |
| | | | **φ2** | -88.9 | |
| | | | **ψ2** | -38.4 | |
| | | | **φ3** | -73.1 | |
| | | | **ψ3** | -53.0 | |
| | | | **φ4** | -112.9 | |
| | | | **ψ4** | -53.0 | |
| | | | **φ5** | -107.4 | |
| | | | | | |
| 15 | APWAD | 10 | **ψ1** | 124.6 | 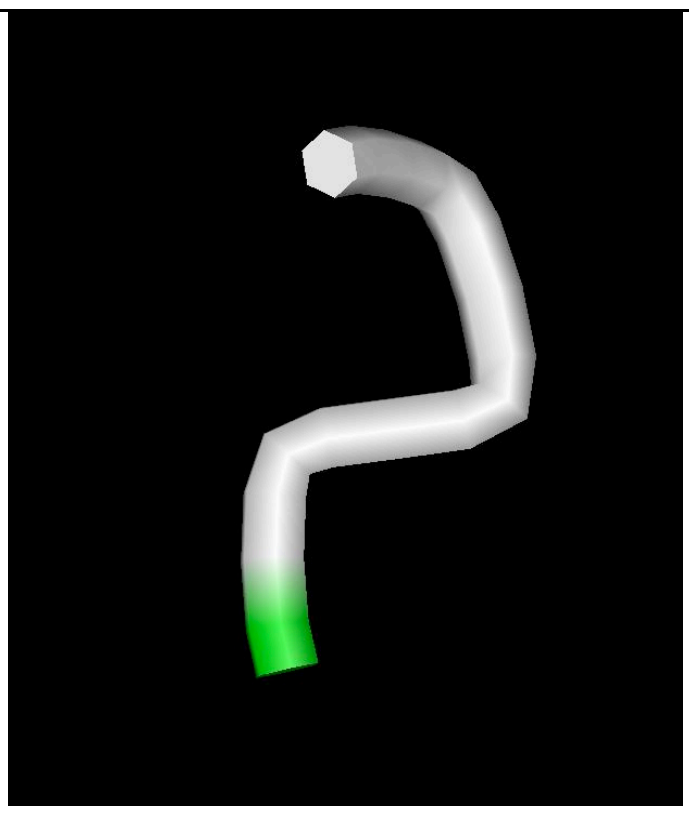 |
| | | | **φ2** | -67.0 | |
| | | | **ψ2** | 108.6 | |
| | | | **φ3** | -97.8 | |
| | | | **ψ3** | 27.0 | |
| | | | **φ4** | -143.7 | |
| | | | **ψ4** | 27.0 | |
| | | | **φ5** | -141.1 | |
| | | | | | |
| 16 | ADRQA | 10 | **ψ1** | -58.1 | 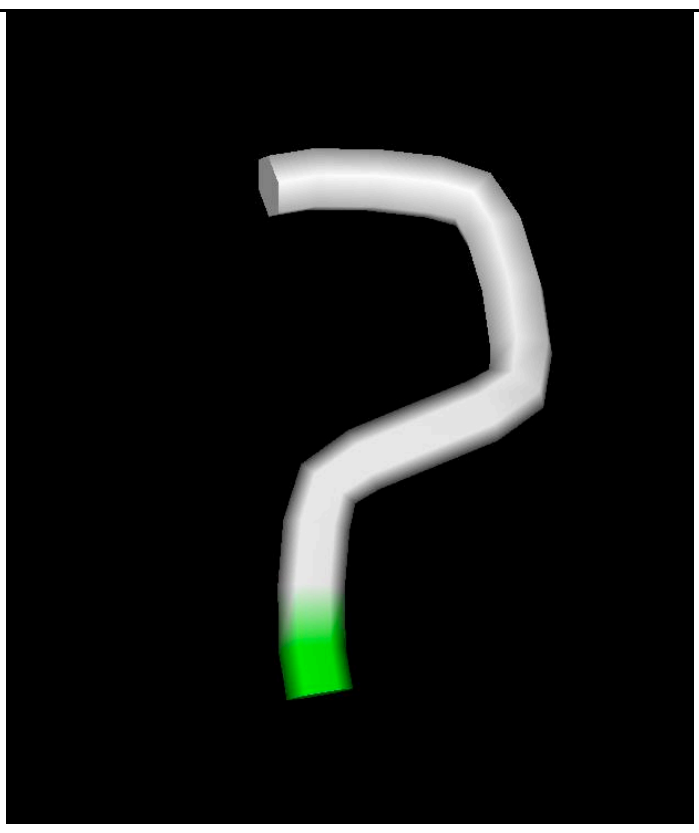 |
| | | | **φ2** | -83.1 | |
| | | | **ψ2** | 145.7 | |
| | | | **φ3** | -130.4 | |
| | | | **ψ3** | -40.0 | |
| | | | **φ4** | -100.3 | |
| | | | **ψ4** | -40.0 | |
| | | | **φ5** | -109.5 | |
| | | | | | |
| 17 | PADTA | 9 | **ψ1** | -47.0 | 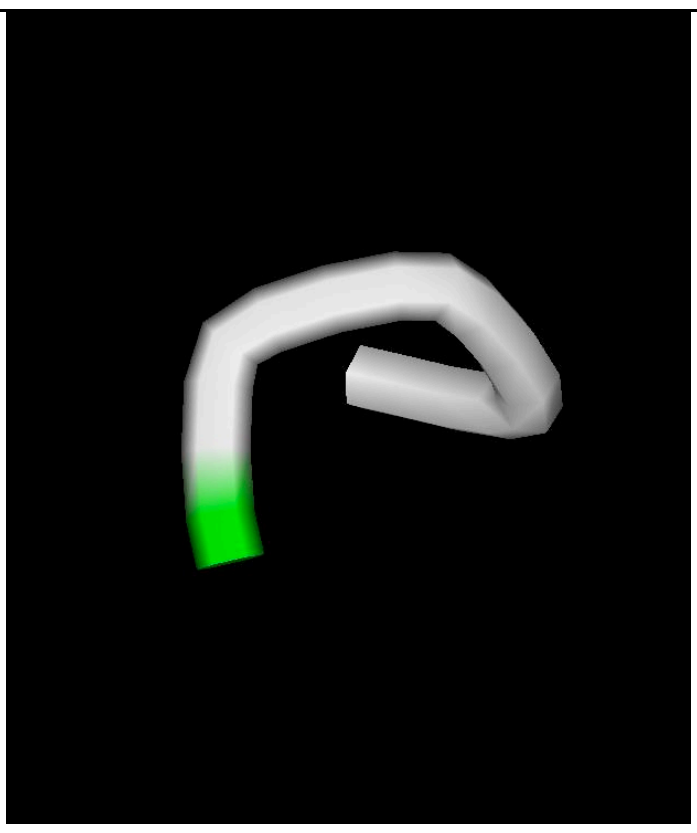 |
| | | | **φ2** | -147.4 | |
| | | | **ψ2** | 49.2 | |
| | | | **φ3** | -149.6 | |
| | | | **ψ3** | -27.8 | |
| | | | **φ4** | -99.5 | |
| | | | **ψ4** | -27.8 | |
| | | | **φ5** | -124.5 | |
| | | | | | |
| 18 | KANAR | 9 | **ψ1** | 13.7 | 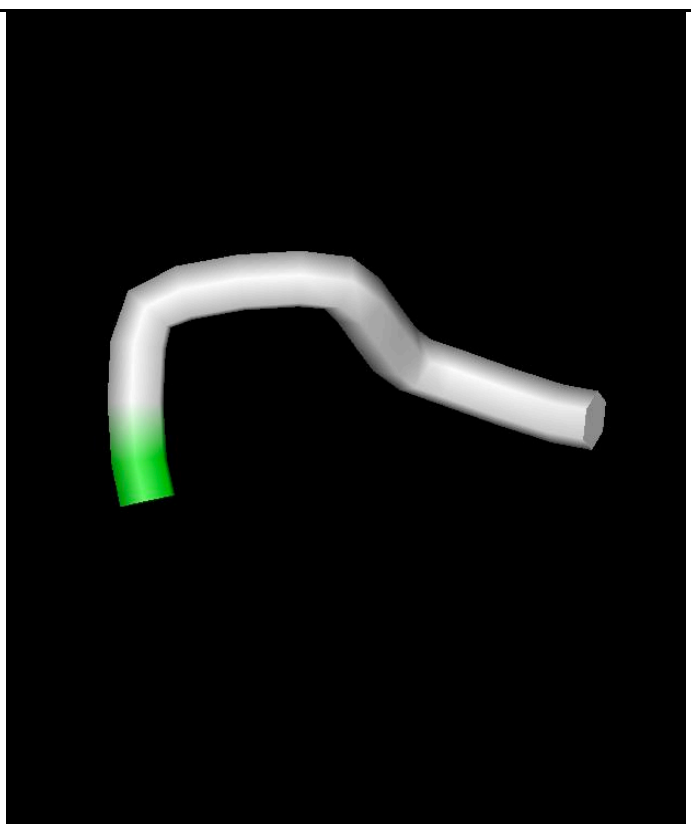 |
| | | | **φ2** | -92.3 | |
| | | | **ψ2** | -45.7 | |
| | | | **φ3** | -86.9 | |
| | | | **ψ3** | 116.7 | |
| | | | **φ4** | -92.4 | |
| | | | **ψ4** | 116.7 | |
| | | | **φ5** | -98.4 | |
| | | | | | |
| 19 | APGTA | 6 | **ψ1** | 118.8 | 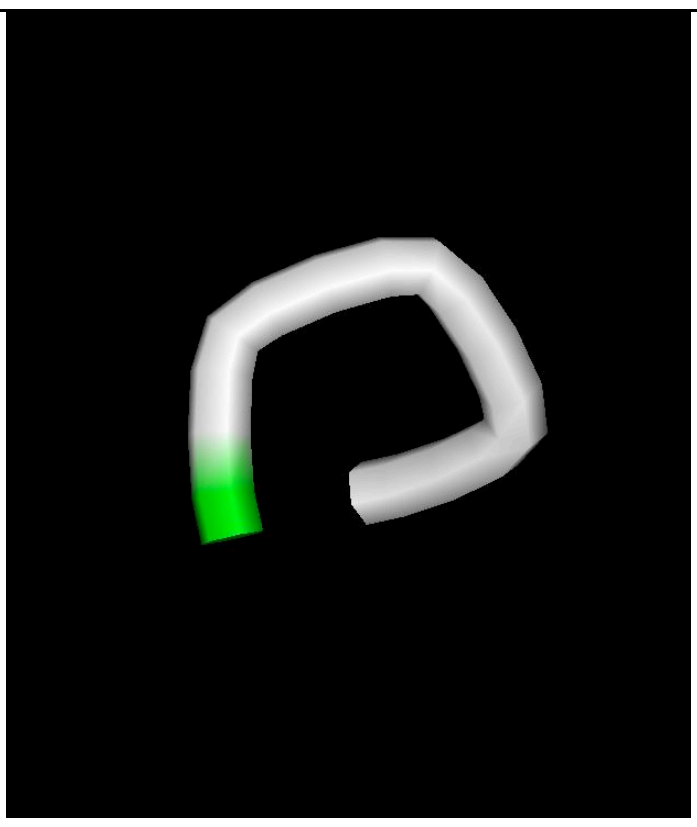 |
| | | | **φ2** | -68.6 | |
| | | | **ψ2** | 124.6 | |
| | | | **φ3** | 126.0 | |
| | | | **ψ3** | -50.7 | |
| | | | **φ4** | -99.6 | |
| | | | **ψ4** | -50.7 | |
| | | | **φ5** | -137.1 | |
| | | | | | |
| 20 | PDNAA | 4 | **ψ1** | 58.5 | 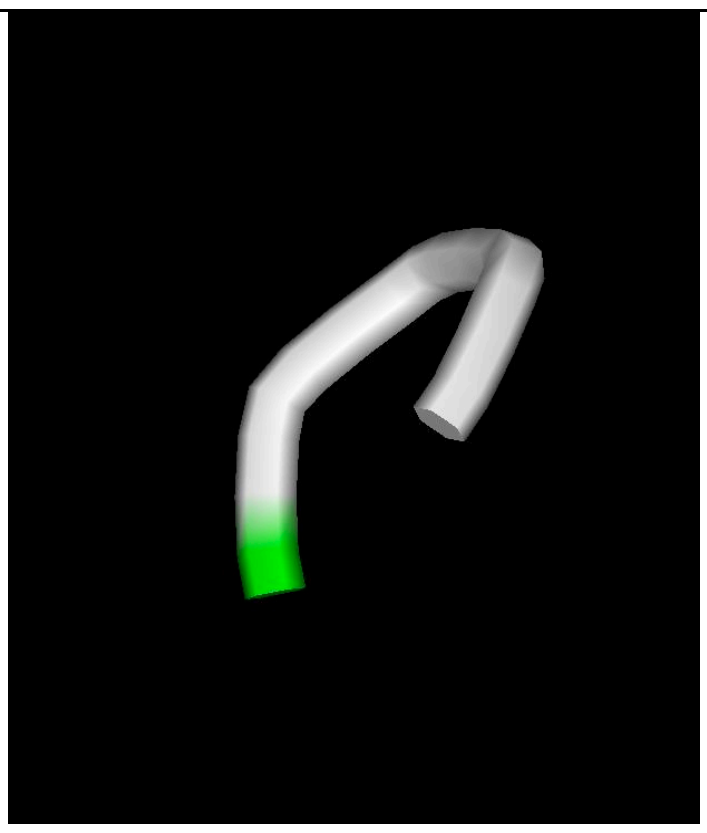 |
| | | | **φ2** | -154.5 | |
| | | | **ψ2** | -112.8 | |
| | | | **φ3** | -129.7 | |
| | | | **ψ3** | -48.4 | |
| | | | **φ4** | -98.2 | |
| | | | **ψ4** | -48.4 | |
| | | | **φ5** | -114.9 | |
| | | | | | |
| 21 | EAHAW | 3 | **ψ1** | -62.1 | 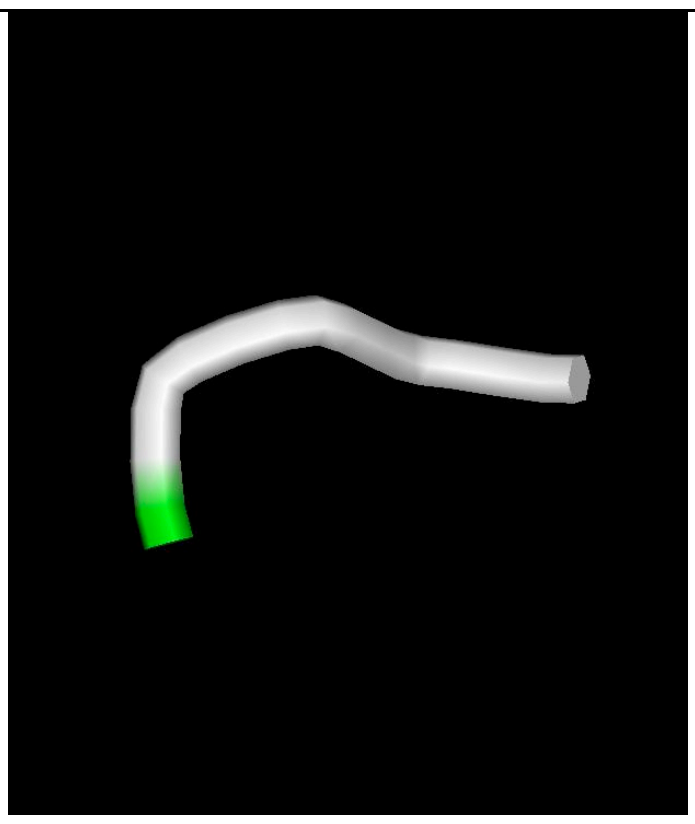 |
| | | | **φ2** | -154.2 | |
| | | | **ψ2** | -44.7 | |
| | | | **φ3** | -97.7 | |
| | | | **ψ3** | 127.6.2 | |
| | | | **φ4** | -97.1 | |
| | | | **ψ4** | 127.6 | |
| | | | **φ5** | -114.5 | |
| | | | | | |
| 22 | RAEAW | 3 | **ψ1** | 112.0 | 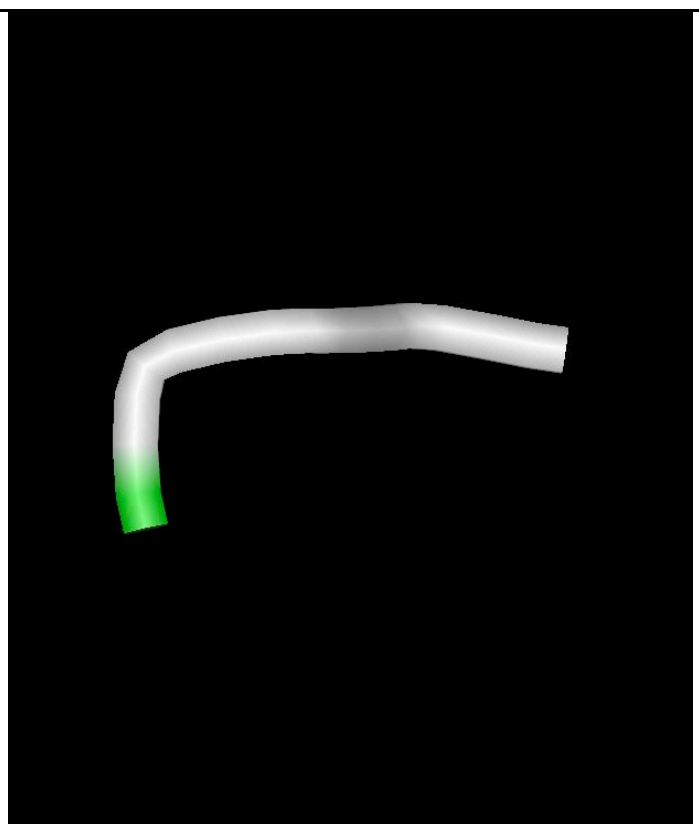 |
| | | | **φ2** | -141.4 | |
| | | | **ψ2** | -20.4 | |
| | | | **φ3** | -79.0 | |
| | | | **ψ3** | 173.2 | |
| | | | **φ4** | -138.8 | |
| | | | **ψ4** | 173.2 | |
| | | | **φ5** | -115.9 | |
| | | | | | |
| 23 | NDIAA | 3 | **ψ1** | -50.2 | 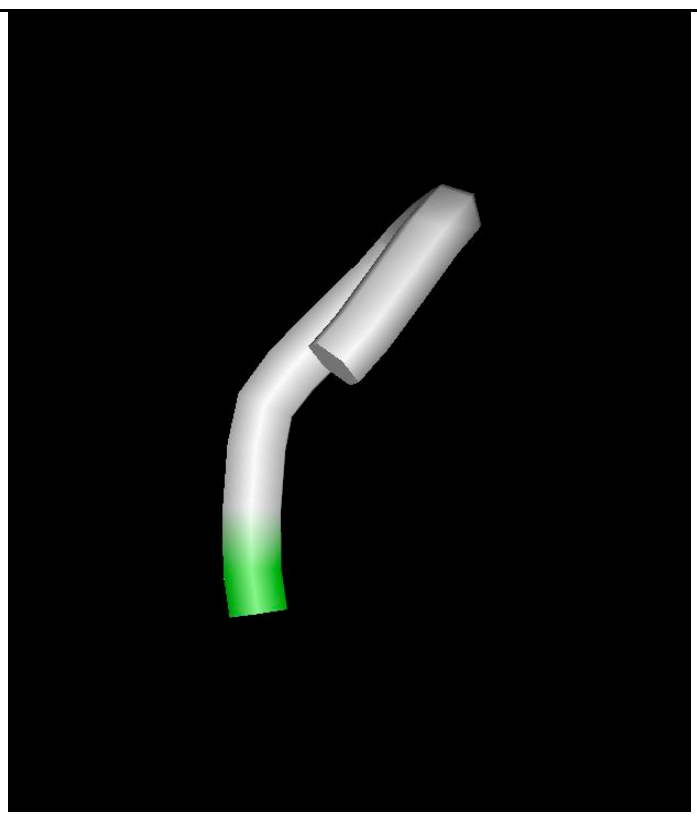 |
| | | | **φ2** | -147.3 | |
| | | | **ψ2** | -142.6 | |
| | | | **φ3** | -113.4 | |
| | | | **ψ3** | -44.2 | |
| | | | **φ4** | -110.4 | |
| | | | **ψ4** | -44.2 | |
| | | | **φ5** | -117.3 | |
| | | | | | |
| 24 | APPAR | 3 | **ψ1** | 114.7 | 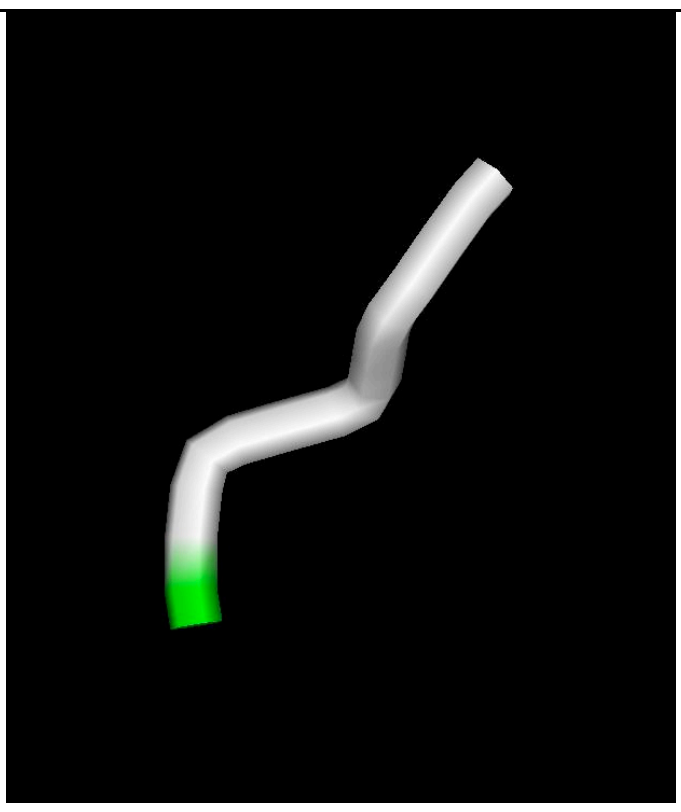 |
| | | | **φ2** | -71.9 | |
| | | | **ψ2** | 135.2 | |
| | | | **φ3** | -70.1 | |
| | | | **ψ3** | 123.4 | |
| | | | **φ4** | -94.8 | |
| | | | **ψ4** | 123.4 | |
| | | | **φ5** | -143.1 | |
| | | | | | |
| 25 | AACDD | 2 | **ψ1** | 80.2 | 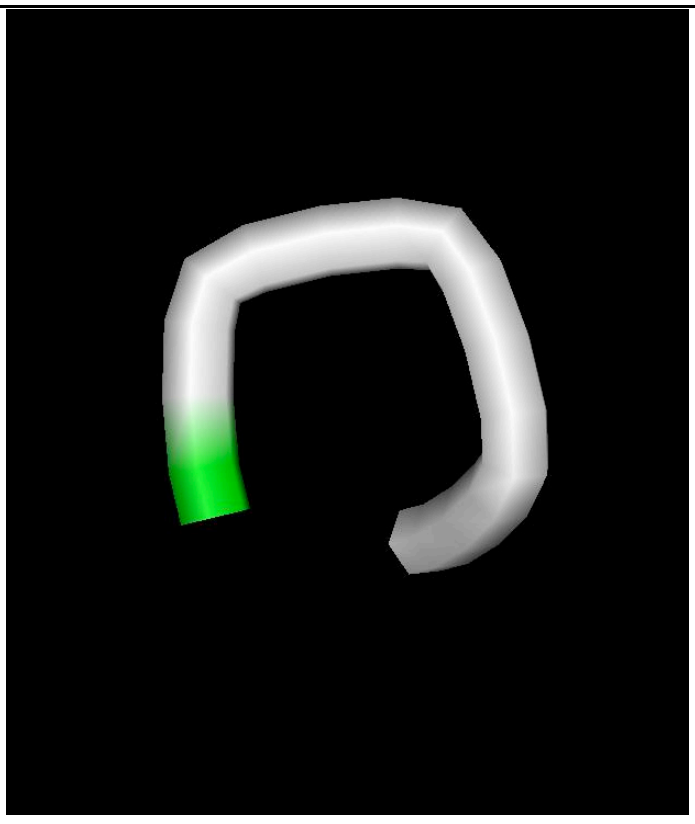 |
| | | | **φ2** | -92.1 | |
| | | | **ψ2** | -48.9 | |
| | | | **φ3** | -128.3 | |
| | | | **ψ3** | 5.4 | |
| | | | **φ4** | -114.2 | |
| | | | **ψ4** | 5.4 | |
| | | | **φ5** | -95.1 | |
| | | | | | |
| 26 | AEHRA | 2 | **ψ1** | 97.8 | 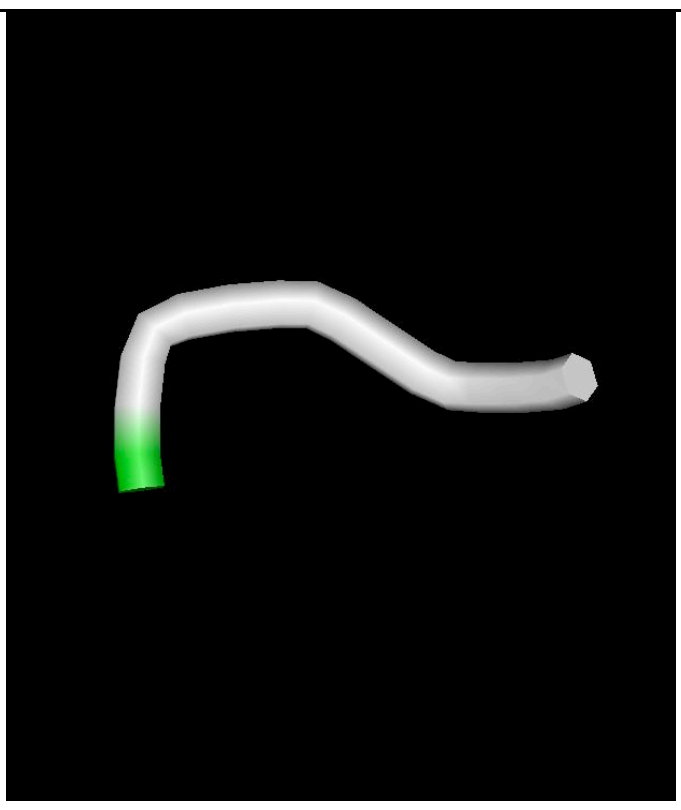 |
| | | | **φ2** | -114.9 | |
| | | | **ψ2** | -48.9 | |
| | | | **φ3** | -142.1 | |
| | | | **ψ3** | 170.7 | |
| | | | **φ4** | -101.9 | |
| | | | **ψ4** | 170.7 | |
| | | | **φ5** | -103.1 | |
| | | | | | |
| 27 | APEAE | 2 | **ψ1** | 96.9 | 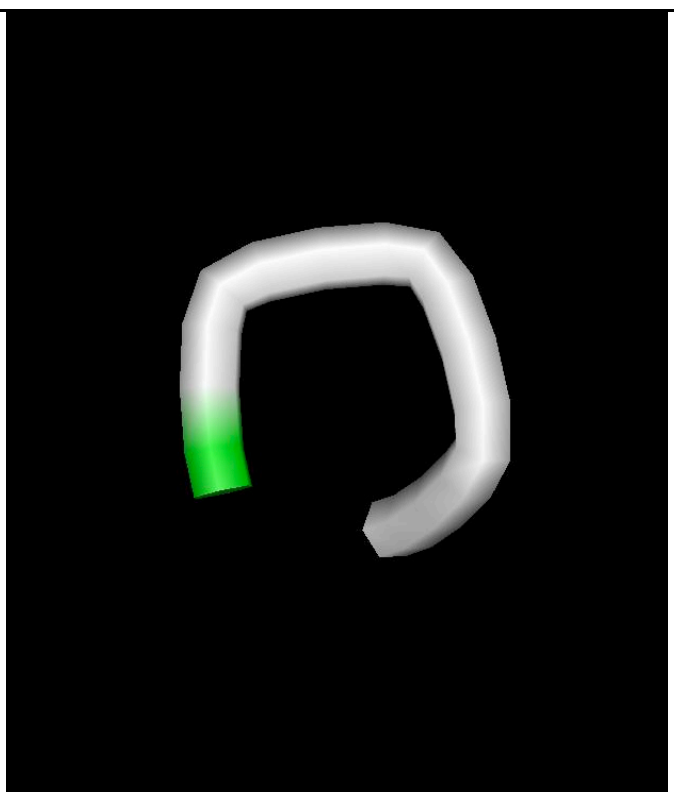 |
| | | | **φ2** | -70.5 | |
| | | | **ψ2** | -0.8 | |
| | | | **φ3** | -75.7 | |
| | | | **ψ3** | 11.5 | |
| | | | **φ4** | -119.6 | |
| | | | **ψ4** | 11.5 | |
| | | | **φ5** | -96.9 | |
| | | | | | |
| 28 | DARGA | 2 | **ψ1** | 143.1 | 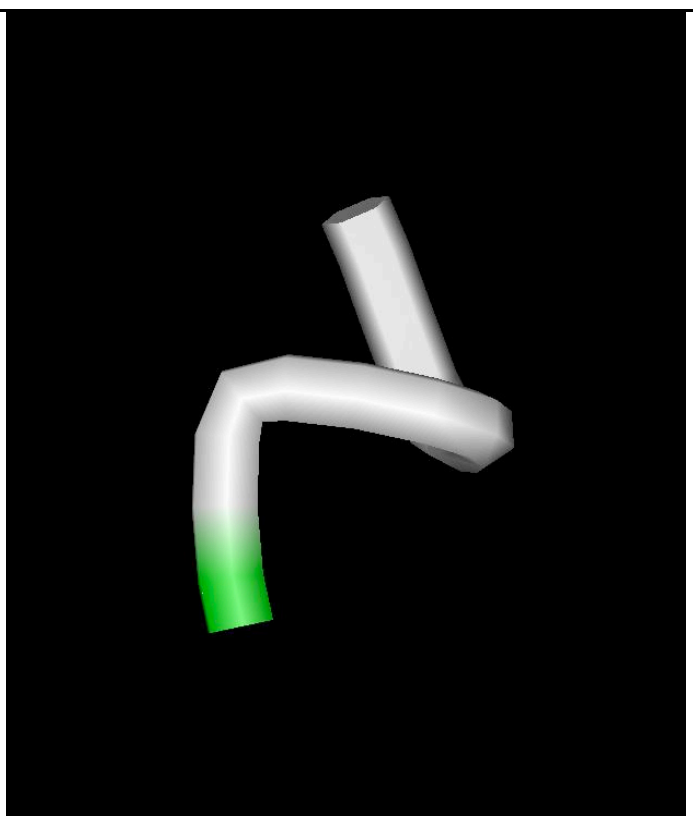 |
| | | | **φ2** | -93.2 | |
| | | | **ψ2** | -14.7 | |
| | | | **φ3** | -95.1 | |
| | | | **ψ3** | -62.8 | |
| | | | **φ4** | 112.0 | |
| | | | **ψ4** | -62.8 | |
| | | | **φ5** | -140.3 | |
| | | | | | |
| 29 | EKDAA | 1 | **ψ1** | -175.8 | 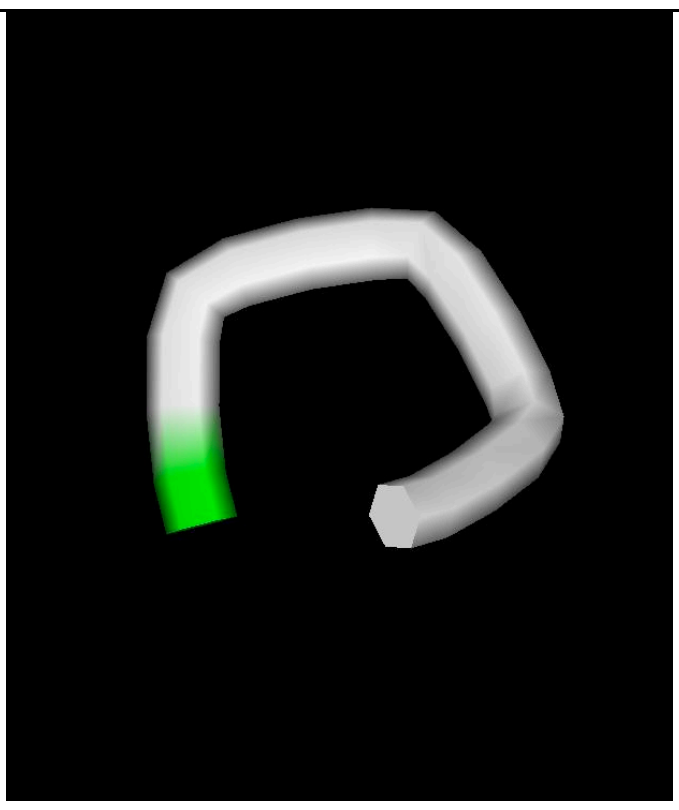 |
| | | | **φ2** | -108.2 | |
| | | | **ψ2** | -41.2 | |
| | | | **φ3** | -136.2 | |
| | | | **ψ3** | -43.9 | |
| | | | **φ4** | -151.5 | |
| | | | **ψ4** | -43.9 | |
| | | | **φ5** | -149.9 | |
| | | | | | |
| 30 | PAGAP | 1 | **ψ1** | -112.4 | 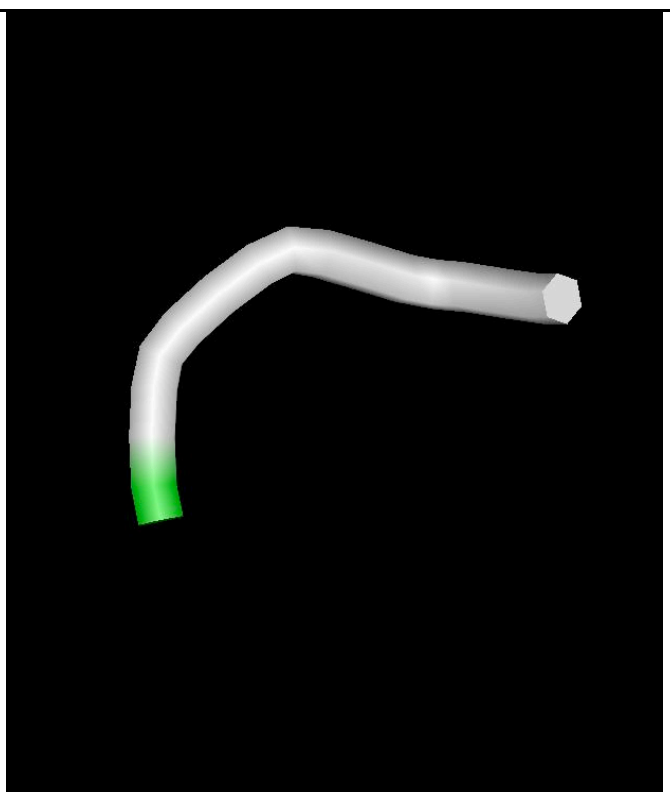 |
| | | | **φ2** | -120.6 | |
| | | | **ψ2** | 146.4 | |
| | | | **φ3** | 85.8 | |
| | | | **ψ3** | 123.1 | |
| | | | **φ4** | -95.0 | |
| | | | **ψ4** | 123.1 | |
| | | | **φ5** | -78.1 | |
| | | | | | |
| 31 | KNGAA | 1 | **ψ1** | -47.7 | 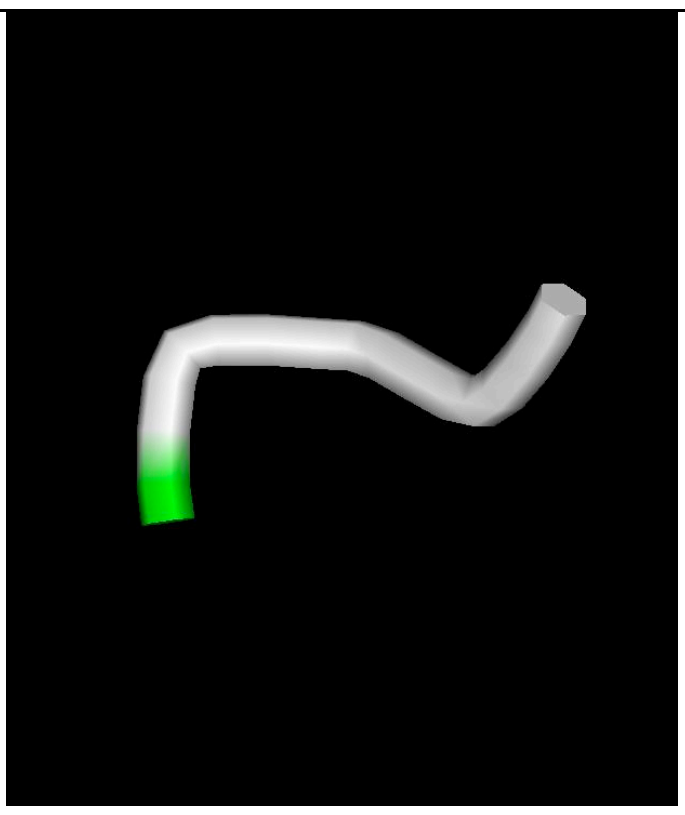 |
| | | | **φ2** | -122.5 | |
| | | | **ψ2** | 22.9 | |
| | | | **φ3** | 112.4 | |
| | | | **ψ3** | 145.2 | |
| | | | **φ4** | -97.3 | |
| | | | **ψ4** | 145.2 | |
| | | | **φ5** | -103.2 | |
| 32 | APPAP | 1 | ψ1 | 116.7 | 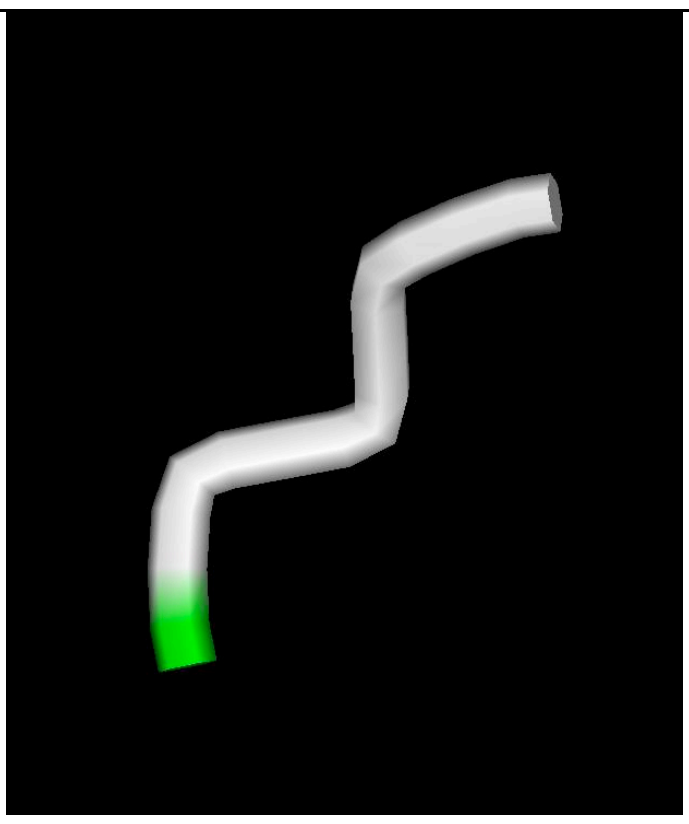 |
| | | | φ2 | -77.3 | |
| | | | ψ2 | 113.3 | |
| | | | φ3 | -76.6 | |
| | | | ψ3 | 109.6 | |
| | | | φ4 | -129.7 | |
| | | | ψ4 | 109.6 | |
| | | | φ5 | -70.0 | |
| | | | | | |
| 33 | KAKDA | 1 | ψ1 | 127.8 | 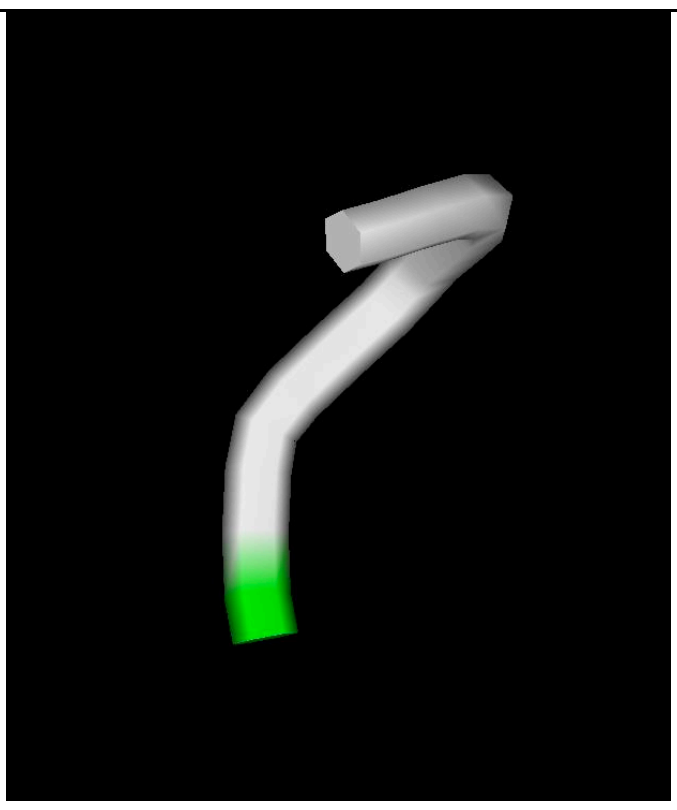 |
| | | | φ2 | -127.0 | |
| | | | ψ2 | -140.3 | |
| | | | φ3 | -103.2 | |
| | | | ψ3 | -85.3 | |
| | | | φ4 | -93.5 | |
| | | | ψ4 | -85.3 | |
| | | | φ5 | -97.8 | |
| | | | | | |
| 34 | EAREA | 1 | ψ1 | 3.4 | 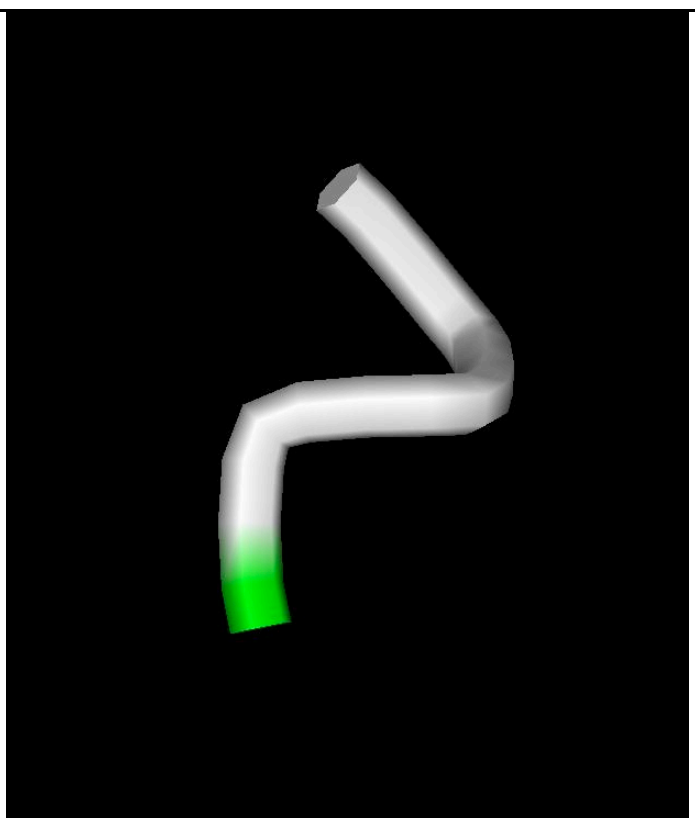 |
| | | | φ2 | -107.6 | |
| | | | ψ2 | 60.1 | |
| | | | φ3 | -111.9 | |
| | | | ψ3 | 11.1 | |
| | | | φ4 | -126.3 | |
| | | | ψ4 | 11.1 | |
| | | | φ5 | -134.5 | |
| | | | | | |
| 35 | WAEYA | 1 | ψ1 | 68.7 | 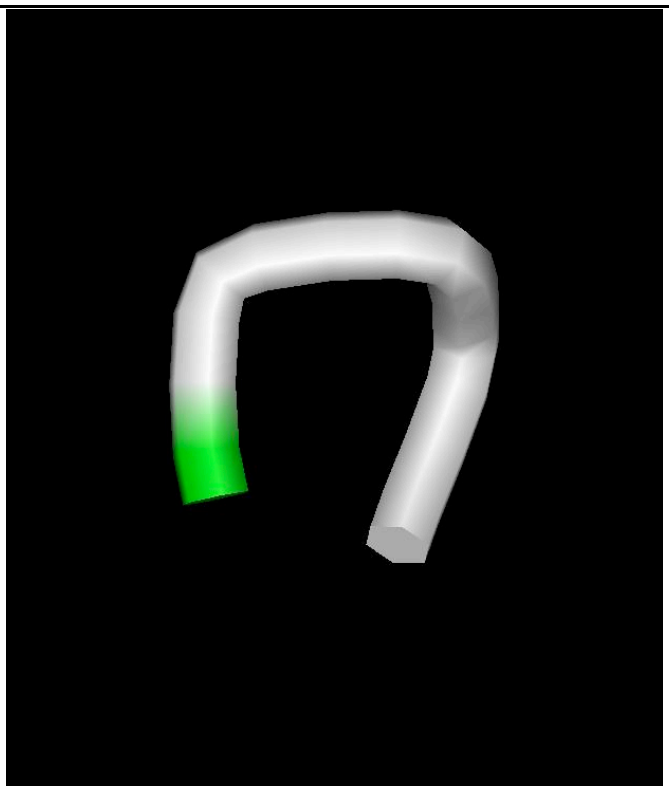 |
| | | | φ2 | -113.4 | |
| | | | ψ2 | -20.3 | |
| | | | φ3 | -92.7 | |
| | | | ψ3 | 42.1 | |
| | | | φ4 | 74.8 | |
| | | | ψ4 | 42.1 | |
| | | | φ5 | -113.5 | |
| 36 | PAKKA | 1 | **ψ1** | -144.7 | 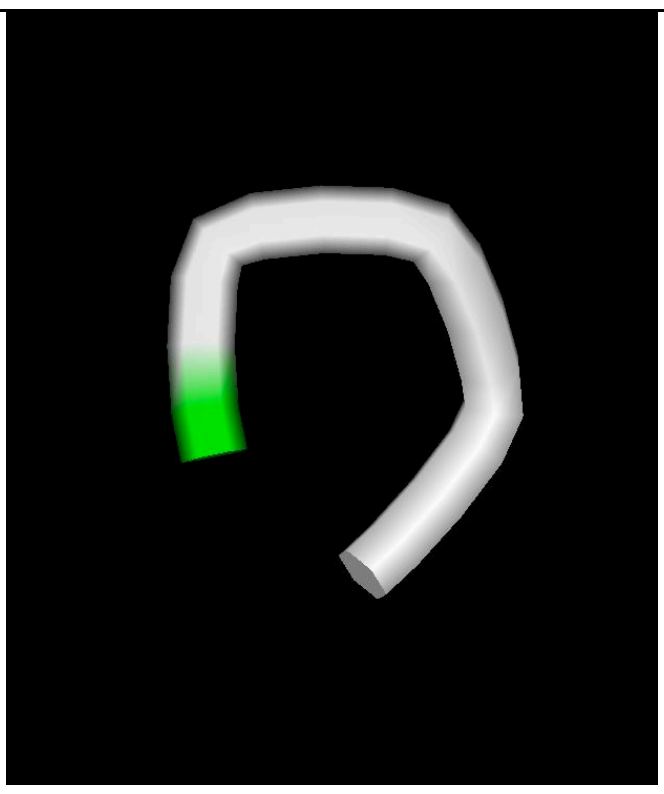 |
| | | | **φ2** | -78.8 | |
| | | | **ψ2** | -39.9 | |
| | | | **φ3** | -74.2 | |
| | | | **ψ3** | -78.6 | |
| | | | **φ4** | -107.8 | |
| | | | **ψ4** | -78.6 | |
| | | | **φ5** | -98.5 | |
| | | | | | |
| 37 | KGYAA | 1 | **ψ1** | -26.6 | 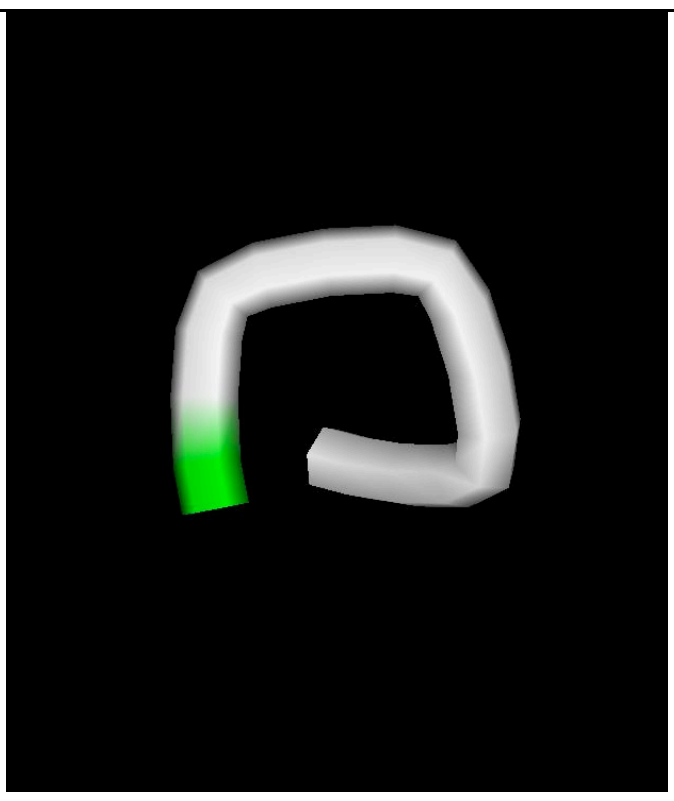 |
| | | | **φ2** | 101.9 | |
| | | | **ψ2** | -69.8 | |
| | | | **φ3** | -98.0 | |
| | | | **ψ3** | -32.3 | |
| | | | **φ4** | -83.2 | |
| | | | **ψ4** | -32.3 | |
| | | | **φ5** | -85.9 | |
| | | | | | |
| 38 | AAWCD | 1 | **ψ1** | -58.8 | 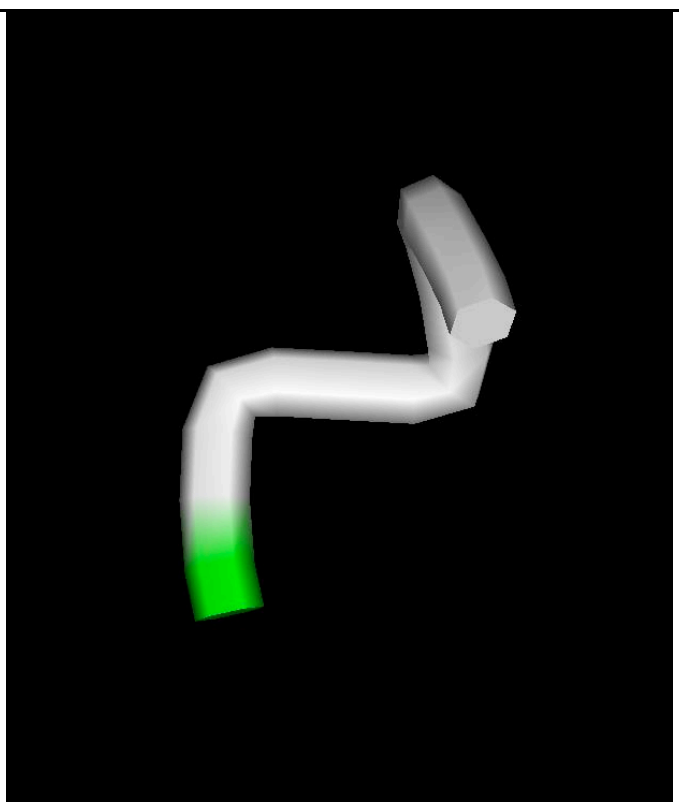 |
| | | | **φ2** | -106.0 | |
| | | | **ψ2** | -10.2 | |
| | | | **φ3** | 61.0 | |
| | | | **ψ3** | 23.9 | |
| | | | **φ4** | -124.1 | |
| | | | **ψ4** | 23.9 | |
| | | | **φ5** | -143.1 | |
| | | | | | |
| 39 | KDKAA | 1 | **ψ1** | -143.6 | 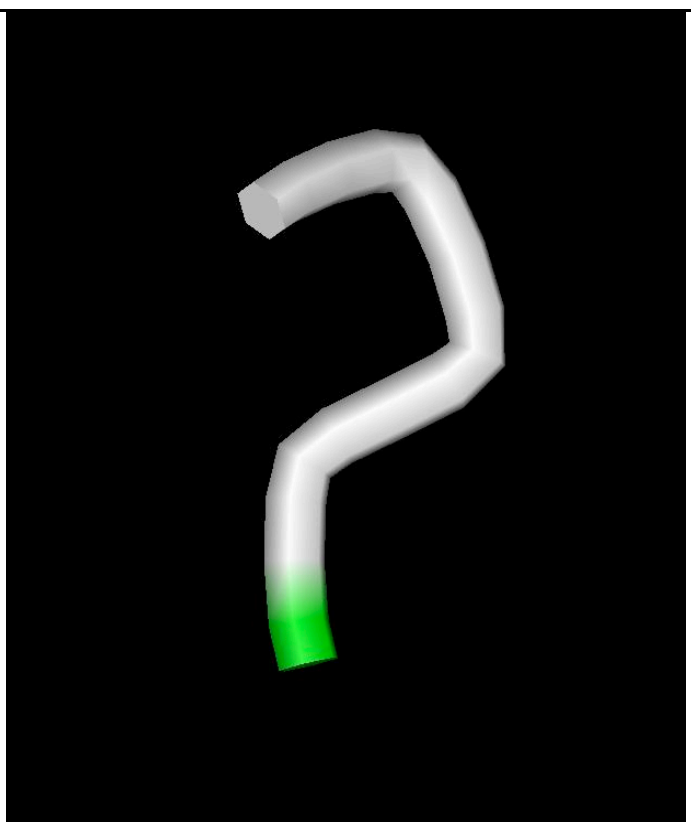 |
| | | | **φ2** | -67.3 | |
| | | | **ψ2** | 162.8 | |
| | | | **φ3** | -125.5 | |
| | | | **ψ3** | -25.9 | |
| | | | **φ4** | -99.2 | |
| | | | **ψ4** | -25.9 | |
| | | | **φ5** | -133.7 | |
| 40 | PAHVA | 1 | ψ1 | 85.2 | 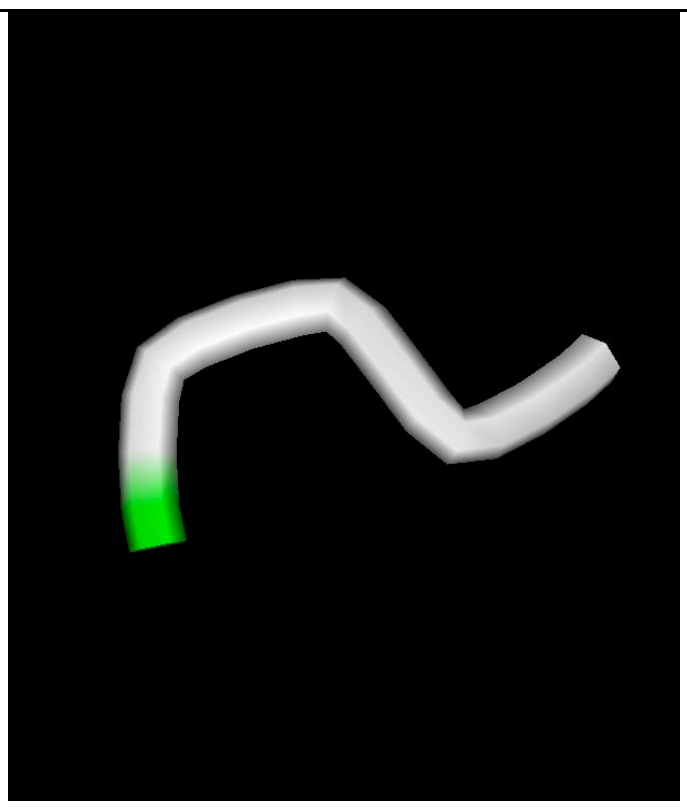 |
| | | | φ2 | -124.2 | |
| | | | ψ2 | -60.3 | |
| | | | φ3 | -111.7 | |
| | | | ψ3 | -70.5 | |
| | | | φ4 | 78.6 | |
| | | | ψ4 | -70.5 | |
| | | | φ5 | -99.7 | |
| | | | | | |
| 41 | RARAY | 1 | ψ1 | -21.4 | 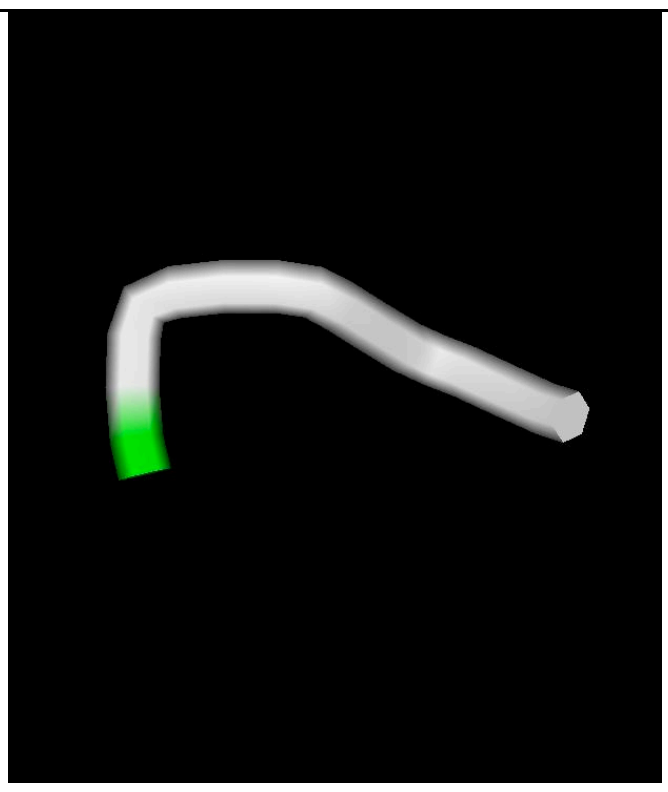 |
| | | | φ2 | -113.1 | |
| | | | ψ2 | 9.8 | |
| | | | φ3 | -145.1 | |
| | | | ψ3 | 127.3 | |
| | | | φ4 | -91.7 | |
| | | | ψ4 | 127.3 | |
| | | | φ5 | -114.7 | |
| | | | | | |
| 42 | AAPKE | 1 | ψ1 | -30.6 | 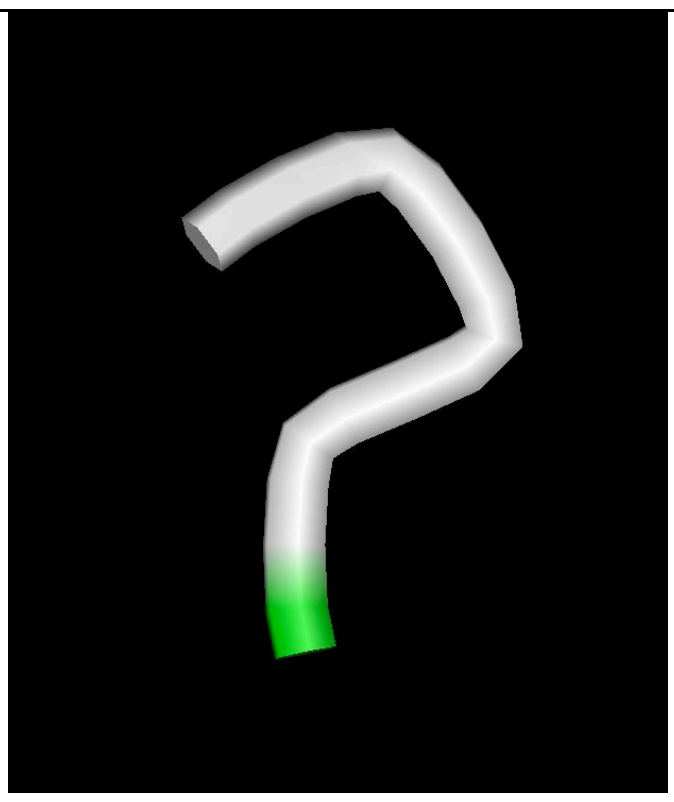 |
| | | | φ2 | -102.3 | |
| | | | ψ2 | 132.6 | |
| | | | φ3 | -86.2 | |
| | | | ψ3 | -5.2 | |
| | | | φ4 | -141.0 | |
| | | | ψ4 | -5.2 | |
| | | | φ5 | -123.4 | |
| | | | | | |
| 43 | APKGA | 1 | ψ1 | 144.7 | 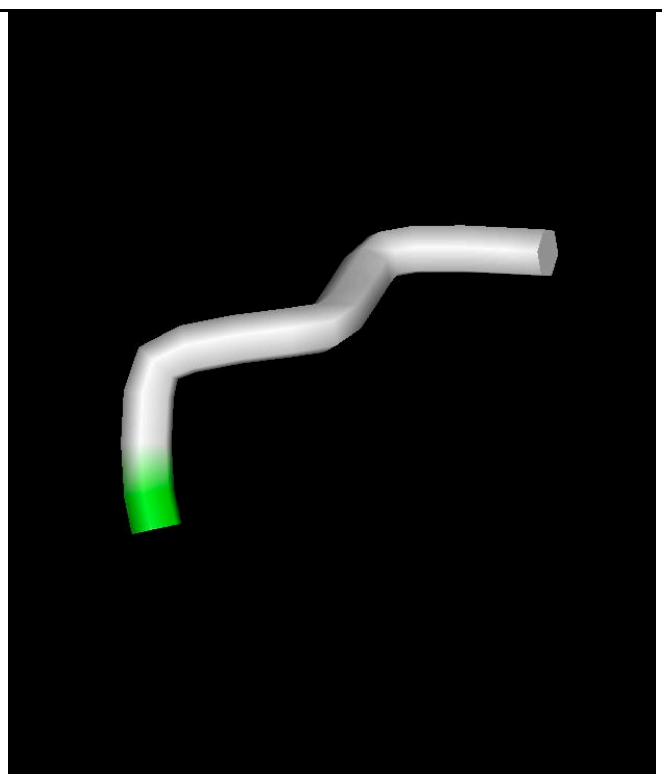 |
| | | | φ2 | -71.3 | |
| | | | ψ2 | 103.6 | |
| | | | φ3 | -129.5 | |
| | | | ψ3 | -76.6 | |
| | | | φ4 | 110.0 | |
| | | | ψ4 | -76.6 | |
| | | | φ5 | -130.1 | |
| 44 | EAKAK | 1 | ψ1 | 166.8 | 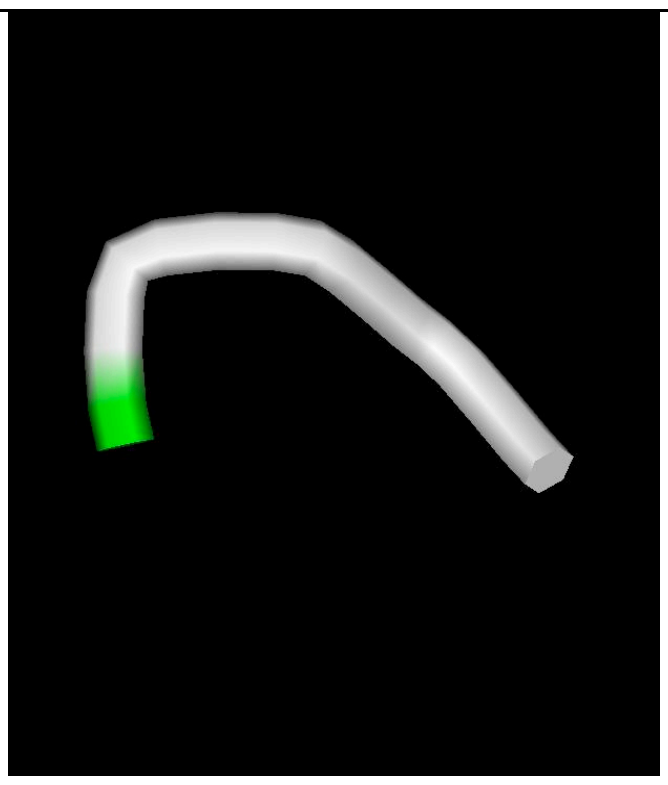 |
| | | | φ2 | -88.1 | |
| | | | ψ2 | -35.4 | |
| | | | φ3 | -103.5 | |
| | | | ψ3 | 157.9 | |
| | | | φ4 | -84.7 | |
| | | | ψ4 | 157.9 | |
| | | | φ5 | -130.1 | |
| | | | | | |
| 45 | FDGAA | | ψ1 | 109.8 | 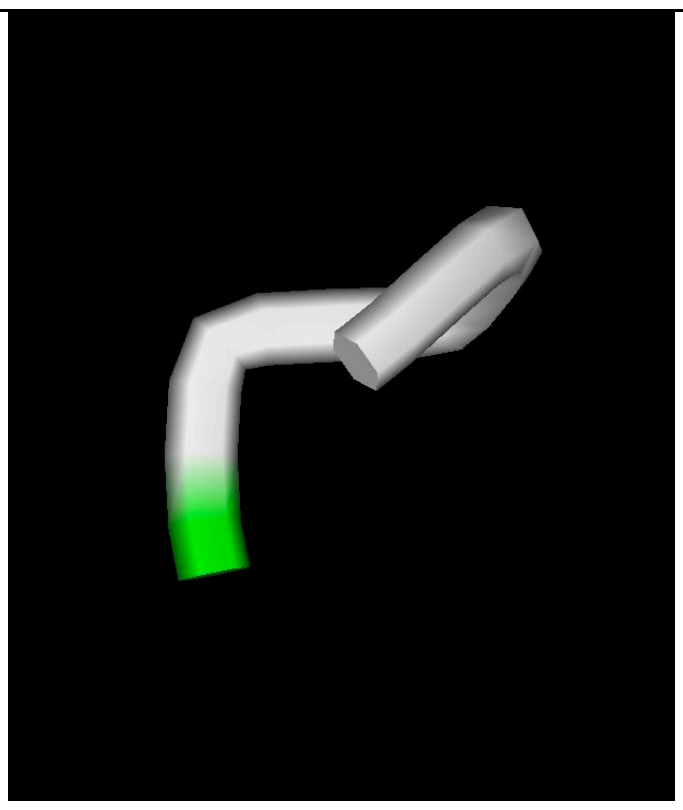 |
| | | | φ2 | -107.1 | |
| | | | ψ2 | -33.3 | |
| | | | φ3 | 140.1 | |
| | | | ψ3 | -50.8 | |
| | | | φ4 | -94.4 | |
| | | | ψ4 | -50.8 | |
| | | | φ5 | -110.5 | |
| | | | | | |
| 46 | WAEAW | 1 | ψ1 | 53.3 | 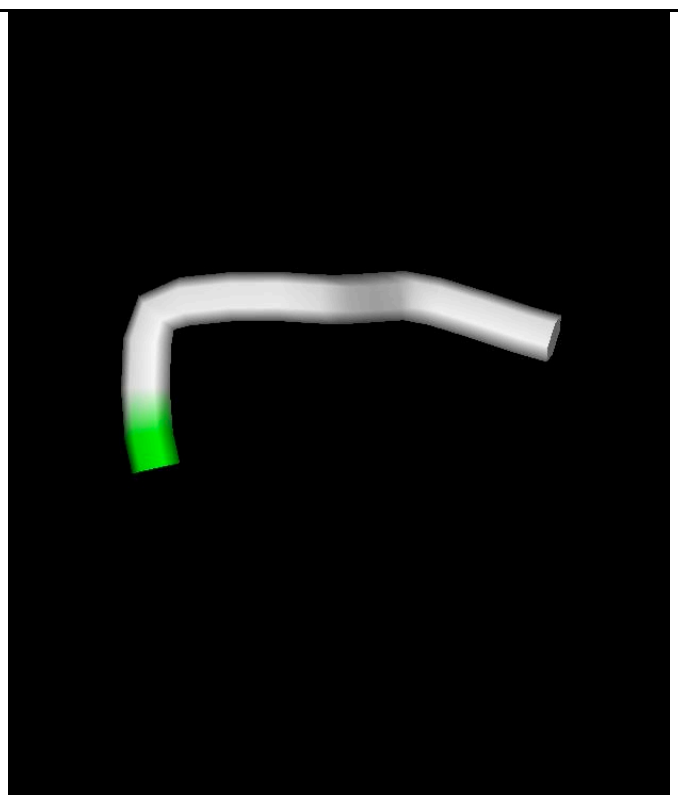 |
| | | | φ2 | -115.8 | |
| | | | ψ2 | -2.0 | |
| | | | φ3 | -98.7 | |
| | | | ψ3 | 143.5 | |
| | | | φ4 | -118.2 | |
| | | | ψ4 | 143.5 | |
| | | | φ5 | -122.0 | |
| | | | | | |
| 47 | PDPAA | 1 | ψ1 | 57.2 | 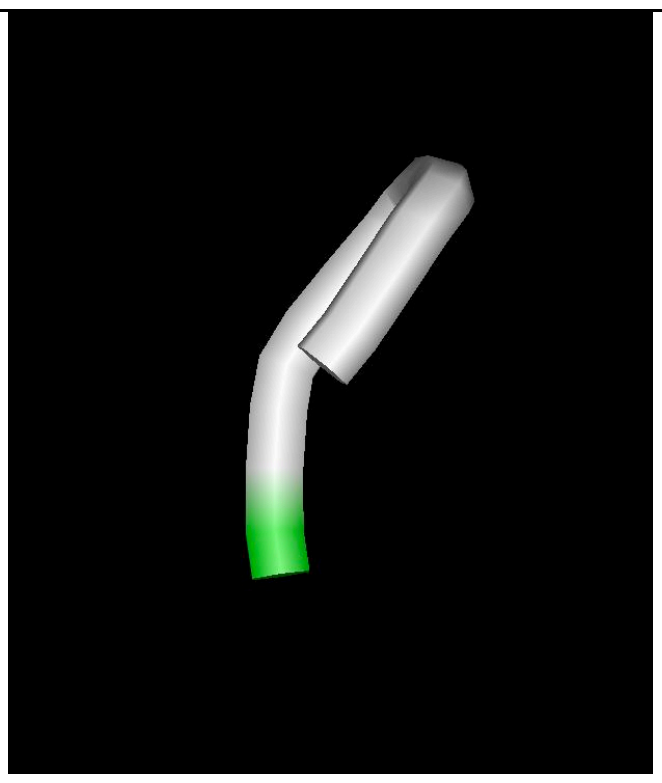 |
| | | | φ2 | -157.6 | |
| | | | ψ2 | -170.3 | |
| | | | φ3 | -71.5 | |
| | | | ψ3 | -39.5 | |
| | | | φ4 | -115.6 | |
| | | | ψ4 | -39.5 | |
| | | | φ5 | -111.4 | |
| 48 | RAMAD | 1 | ψ1 | -46.9 | 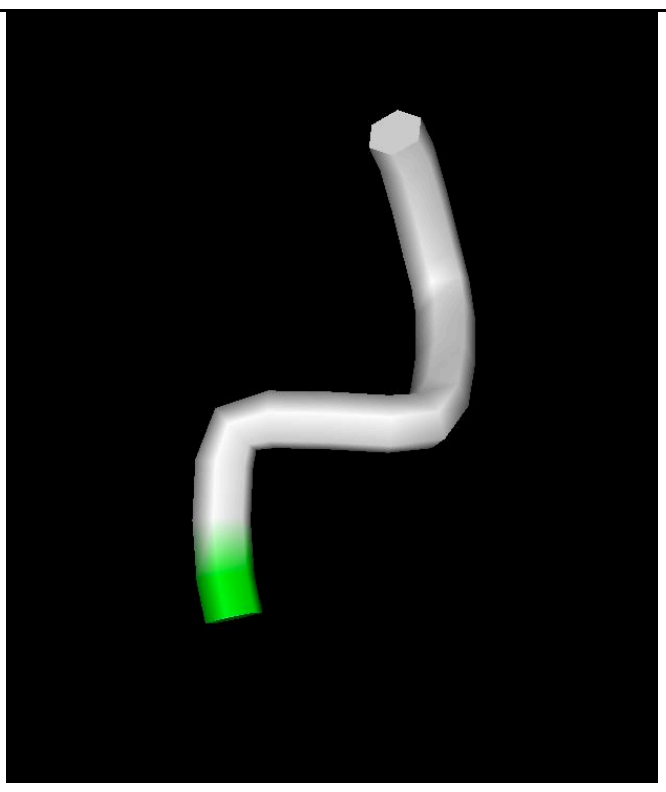 |
| | | | φ2 | 60.7 | |
| | | | ψ2 | 53.7 | |
| | | | φ3 | -115.8 | |
| | | | ψ3 | 35.9 | |
| | | | φ4 | -134.8 | |
| | | | ψ4 | 35.9 | |
| | | | φ5 | -154.7 | |
| | | | | | |
| 49 | KREAA | 1 | ψ1 | 62.5 | 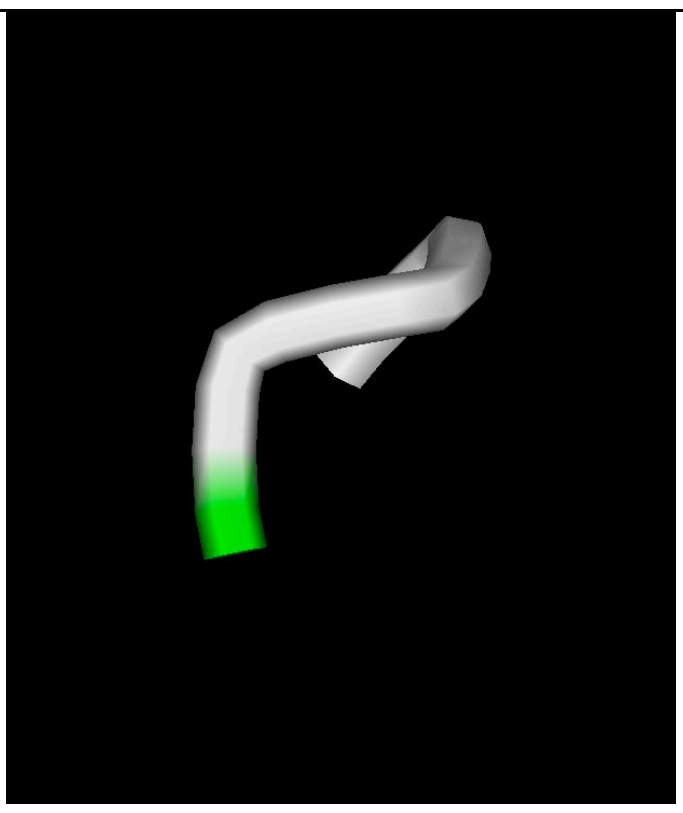 |
| | | | φ2 | -151.6 | |
| | | | ψ2 | 49.4 | |
| | | | φ3 | -93.2 | |
| | | | ψ3 | -49.2 | |
| | | | φ4 | -140.8 | |
| | | | ψ4 | -49.2 | |
| | | | φ5 | -127.8 | |
| | | | | | |
| 50 | PACAK | 1 | ψ1 | -131.2 | 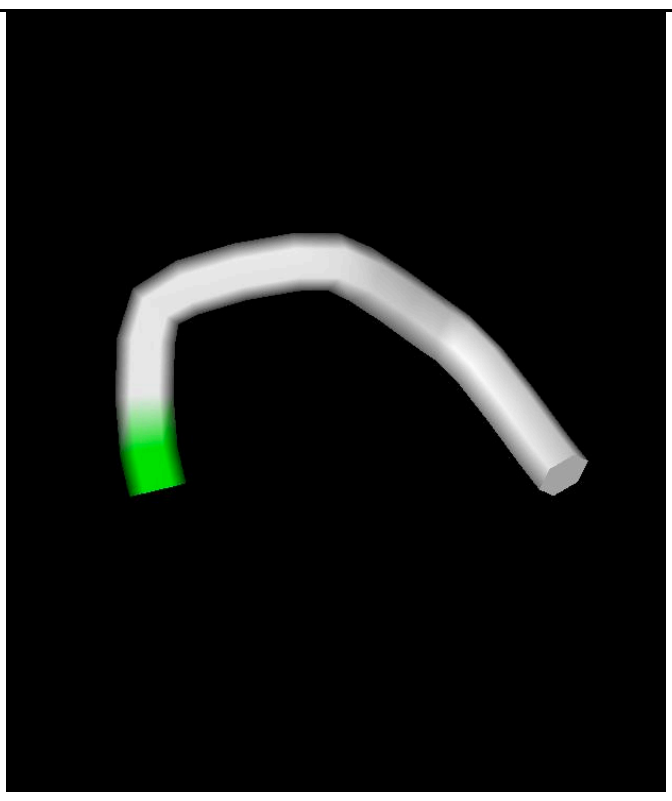 |
| | | | φ2 | -91.2 | |
| | | | ψ2 | -61.3 | |
| | | | φ3 | -76.4 | |
| | | | ψ3 | 156.4 | |
| | | | φ4 | -76.5 | |
| | | | ψ4 | 156.4 | |
| | | | φ5 | -97.0 | |
| | | | | | |
| 51 | DAICA | 1 | ψ1 | 142.1 | 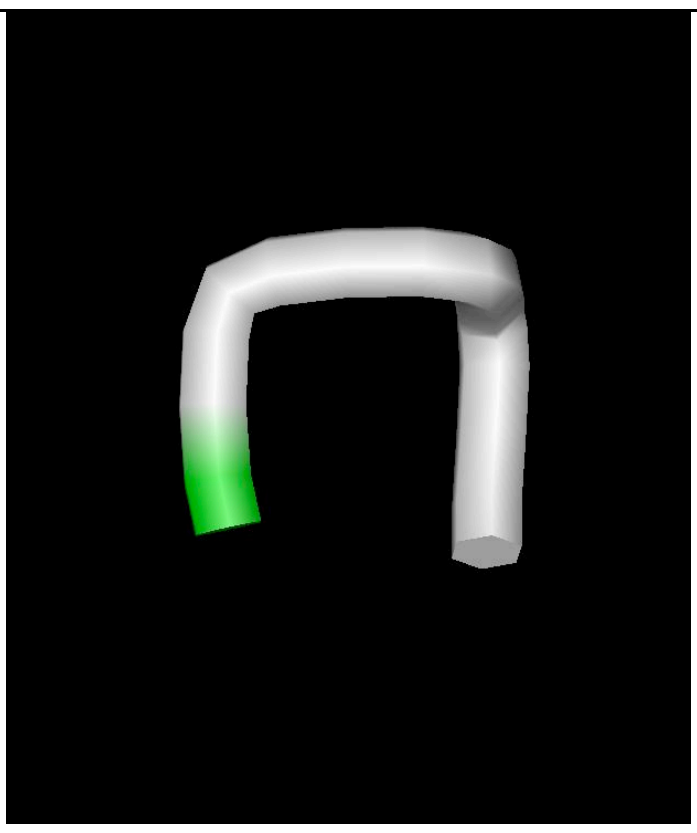 |
| | | | φ2 | -119.1 | |
| | | | ψ2 | -1.3 | |
| | | | φ3 | -99.7 | |
| | | | ψ3 | 44.2 | |
| | | | φ4 | 55.4 | |
| | | | ψ4 | 44.2 | |
| | | | φ5 | -155.1 | |
| | | | | | |
| 52 | RAEAD | 1 | **ψ1** | 119.4. | 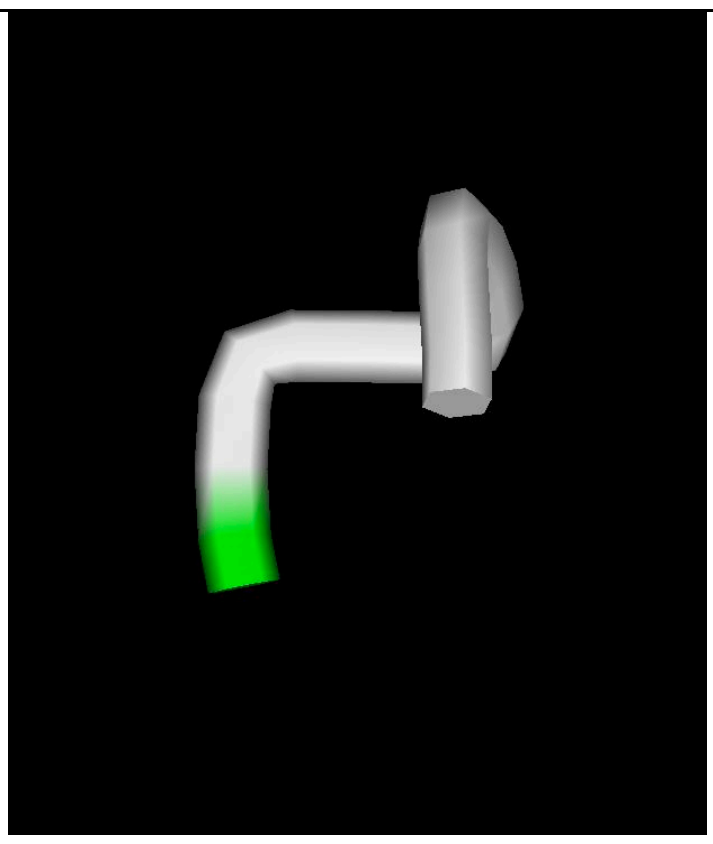 |
| | | | **φ2** | -123.1 | |
| | | | **ψ2** | 5.2 | |
| | | | **φ3** | 65.3 | |
| | | | **ψ3** | 14.6 | |
| | | | **φ4** | -128.0 | |
| | | | **ψ4** | 14.6 | |
| | | | **φ5** | -139.8 | |
| | | | | | |
| 53 | APKKA | 1 | **ψ1** | 134.5 | 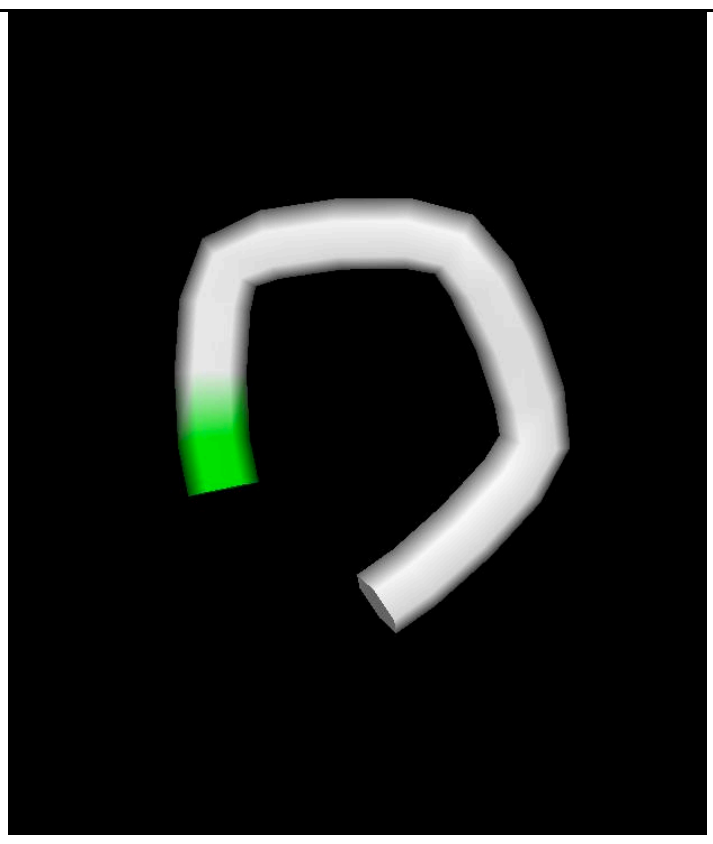 |
| | | | **φ2** | -69.4 | |
| | | | **ψ2** | -55.5 | |
| | | | **φ3** | -69.8 | |
| | | | **ψ3** | -78.3 | |
| | | | **φ4** | -100.0 | |
| | | | **ψ4** | -78.3 | |
| | | | **φ5** | -96.6 | |
| | | | | | |
| 54 | EAGKA | 1 | **ψ1** | 115.7 | 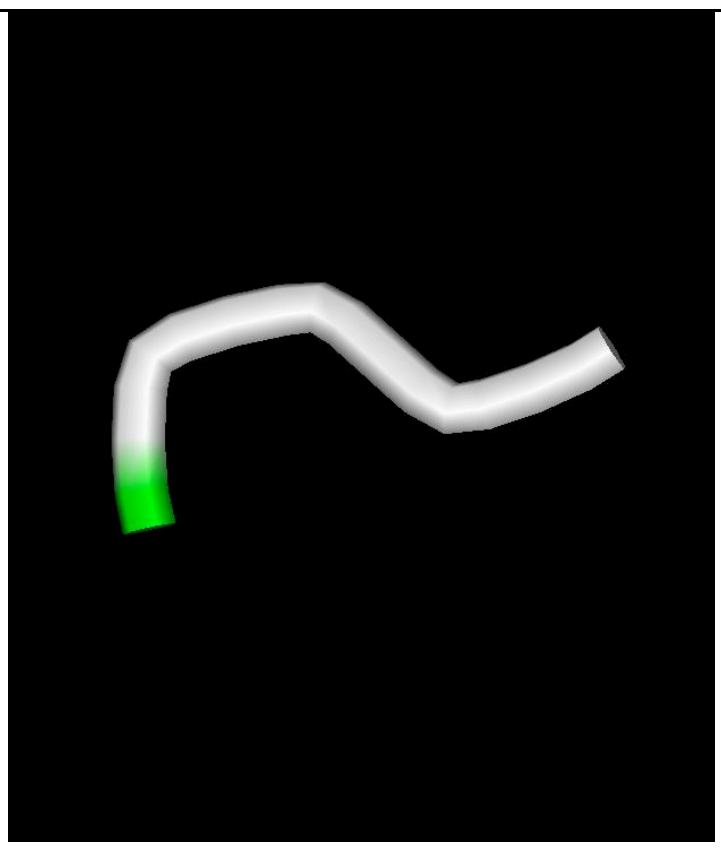 |
| | | | **φ2** | -128.5 | |
| | | | **ψ2** | 74.1 | |
| | | | **φ3** | 98.2 | |
| | | | **ψ3** | 117.0 | |
| | | | **φ4** | -137.0 | |
| | | | **ψ4** | 117.0 | |
| | | | **φ5** | -120.7 | |

**Supplementary Table 3.** The root-mean-square deviation (RMSD) between the three N- or C-terminal C-alpha atoms of rigid pentapeptides and C-alpha atoms of idealized elements of the secondary structure. The smaller the RMSD value, the higher the similarity of the fragment of the rigid pentapeptide to the idealized secondary structure.

| Cluster number | RMSD between N-end of pentapeptide and α-helix | RMSD between N-end of pentapeptide and ↑↓β-structure | RMSD between N-end of pentapeptide and ↑↑β-structure | RMSD between C- end of pentapeptide and α-helix | RMSD between C-end of pentapeptide and ↑↓β-structure | RMSD between C-end of pentapeptide and ↑↑β-structure |
|---|---|---|---|---|---|---|
| 1 | 0.113 | 0.573 | 0.369 | 0.605 | 0.136 | 0.071 |
| 2 | 0.130 | 0.556 | 0.352 | 0.280 | 0.432 | 0.227 |
| 3 | 0.150 | 0.537 | 0.332 | 0.081 | 0.612 | 0.408 |
| 4 | 0.349 | 0.339 | 0.133 | 0.262 | 0.449 | 0.244 |
| 5 | 0.164 | 0.523 | 0.318 | 0.369 | 0.351 | 0.146 |
| 6 | 0.298 | 0.390 | 0.184 | 0.813 | 0.060 | 0.266 |
| 7 | 0.119 | 0.567 | 0.363 | 0.075 | 0.617 | 0.413 |
| 8 | 0.431 | 0.257 | 0.051 | 0.488 | 0.243 | 0.038 |
| 9 | 0.040 | 0.645 | 0.441 | 0.422 | 0.303 | 0.098 |
| 10 | 0.561 | 0.125 | 0.081 | 0.074 | 0.618 | 0.414 |
| 11 | 0.141 | 0.545 | 0.340 | 0.092 | 0.602 | 0.398 |
| 12 | 0.001 | 0.684 | 0.481 | 0.028 | 0.710 | 0.507 |
| 13 | 0.025 | 0.709 | 0.506 | 0.147 | 0.553 | 0.349 |
| 14 | 0.075 | 0.610 | 0.406 | 0.282 | 0.431 | 0.226 |
| 15 | 0.226 | 0.461 | 0.256 | 0.192 | 0.512 | 0.308 |
| 16 | 0.496 | 0.191 | 0.015 | 0.149 | 0.551 | 0.346 |
| 17 | 0.245 | 0.442 | 0.236 | 0.082 | 0.611 | 0.407 |
| 18 | 0.130 | 0.556 | 0.352 | 0.404 | 0.320 | 0.114 |
| 19 | 0.330 | 0.357 | 0.151 | 0.209 | 0.497 | 0.292 |
| 20 | 0.679 | 0.006 | 0.201 | 0.188 | 0.515 | 0.311 |
| 21 | 0.312 | 0.375 | 0.169 | 0.496 | 0.236 | 0.030 |
| 22 | 0.200 | 0.487 | 0.282 | 0.885 | 0.128 | 0.334 |
| 23 | 0.803 | 0.122 | 0.328 | 0.219 | 0.488 | 0.283 |
| 24 | 0.402 | 0.285 | 0.079 | 0.459 | 0.270 | 0.064 |
| 25 | 0.147 | 0.539 | 0.334 | 0.055 | 0.635 | 0.431 |
| 26 | 0.241 | 0.446 | 0.241 | 0.731 | 0.018 | 0.188 |
| 27 | 0.153 | 0.834 | 0.632 | 0.077 | 0.616 | 0.412 |
| 28 | 0.005 | 0.689 | 0.486 | 0.163 | 0.538 | 0.334 |
| 29 | 0.175 | 0.511 | 0.306 | 0.337 | 0.381 | 0.176 |
| 30 | 0.647 | 0.038 | 0.168 | 0.458 | 0.271 | 0.065 |
| 31 | 0.085 | 0.600 | 0.396 | 0.602 | 0.138 | 0.068 |
| 32 | 0.290 | 0.398 | 0.192 | 0.530 | 0.204 | 0.002 |
| 33 | 0.754 | 0.071 | 0.277 | 0.411 | 0.314 | 0.108 |
| 34 | 0.116 | 0.570 | 0.366 | 0.108 | 0.588 | 0.384 |
| 35 | 0.105 | 0.580 | 0.376 | 0.030 | 0.658 | 0.454 |
| 36 | 0.035 | 0.650 | 0.446 | 0.427 | 0.299 | 0.093 |
| 37 | 0.134 | 0.552 | 0.347 | 0.019 | 0.667 | 0.464 |
| 38 | 0.042 | 0.643 | 0.439 | 0.104 | 0.592 | 0.388 |
| 39 | 0.513 | 0.174 | 0.0327 | 0.072 | 0.620 | 0.417 |
| 40 | 0.334 | 0.354 | 0.148 | 0.038 | 0.651 | 0.447 |
| 41 | 0.041 | 0.644 | 0.440 | 0.470 | 0.259 | 0.053 |
| 42 | 0.501 | 0.186 | 0.0204 | 0.181 | 0.522 | 0.317 |
| 43 | 0.207 | 0.479 | 0.274 | 0.227 | 0.480 | 0.276 |
| 44 | 0.057 | 0.628 | 0.424 | 0.611 | 0.130 | 0.076 |
| 45 | 0.132 | 0.554 | 0.349 | 0.185 | 0.518 | 0.314 |
| 46 | 0.067 | 0.618 | 0.414 | 0.684 | 0.062 | 0.145 |
| 47 | 0.868 | 0.190 | 0.395 | 0.215 | 0.492 | 0.287 |
| 48 | 0.027 | 0.658 | 0.454 | 0.175 | 0.528 | 0.323 |
| 49 | 0.260 | 0.426 | 0.221 | 0.346 | 0.373 | 0.168 |
| 50 | 0.216 | 0.471 | 0.266 | 0.571 | 0.166 | 0.040 |
| 51 | 0.080 | 0.605 | 0.401 | 0.060 | 0.738 | 0.535 |
| 52 | 0.088 | 0.597 | 0.393 | 0.115 | 0.581 | 0.377 |
| 53 | 0.079 | 0.606 | 0.402 | 0.3920 | 0.331 | 0.125 |
| 54 | 0.278 | 0.409 | 0.204 | 0.610 | 0.131 | 0.075 |